\documentclass[twocolumn]{aastex631}
\usepackage{graphicx}

\usepackage{multirow}
\usepackage{colortbl}
\usepackage{amsmath,amssymb}
\usepackage{txfonts}
\usepackage{pifont}
\newcommand{\cmark}{\ding{51}}%
\usepackage{subfigure}

\definecolor{myviolet}{rgb}{0.933, 0.510, 0.933}
\definecolor{myblue}{rgb}{0.0, 0.0, 1.0}
\definecolor{mypurple}{rgb}{0.214, 0.117, 0.4}
\definecolor{mylime}{rgb}{0.77, 0.55, 0.14}
\definecolor{myred}{rgb}{0.73, 0.01, 0.}
\definecolor{mygrey}{rgb}{0.5, 0.5, 0.5}
\definecolor{mybrown}{rgb}{0.35, 0.095, 0.0}

\submitjournal{AJ}

\begin{document}

\title{Spectroscopically resolved partial phase curve of the rapid heating and cooling of the highly-eccentric Hot Jupiter HAT-P-2b with WFC3}

\correspondingauthor{Jean-Michel Désert}
\email{j.m.l.b.desert@uva.nl}

\author[0000-0002-0373-1517]{Bob Jacobs}
\affiliation{Anton Pannekoek Institute for Astronomy, University of Amsterdam,
Science Park 904, 1098 XH,
Amsterdam, the Netherlands}
\affiliation{Department of Astrophysics/IMAPP, Radboud University, PO Box 9010, 6500 GL Nijmegen, The Netherlands}
\author[0000-0002-0875-8401]{Jean-Michel D\'esert}
\affiliation{Anton Pannekoek Institute for Astronomy, University of Amsterdam,
Science Park 904, 1098 XH,
Amsterdam, the Netherlands}
\author[0000-0002-8507-1304]{Nikole Lewis}
\affiliation{Department of Astronomy and Carl Sagan Institute, Cornell University,
122 Sciences Drive,
Ithaca, NY 14853, USA}
\author[0000-0002-8211-6538]{Ryan C. Challener}
\affiliation{Department of Astronomy and Carl Sagan Institute, Cornell University,
122 Sciences Drive,
Ithaca, NY 14853, USA}
\author[0000-0002-4321-4581]{L. C. Mayorga}
\affiliation{Johns Hopkins University Applied Physics Laboratory, 11100 Johns Hopkins Rd, Laurel, MD, 20723, USA}

\author[0000-0002-7564-6047]{Zo{\"e}\ L. de Beurs}
\affiliation{Department of Earth, Atmospheric and Planetary Sciences, Massachusetts Institute of Technology, Cambridge, MA 02139, USA}
\affiliation{NSF Graduate Research Fellow, MIT Presidential Fellow, MIT Collamore-Rogers Fellow}
\author{Vivien Parmentier}
\affiliation{Laboratoire Lagrange, Observatoire de la Côte d’Azur, Universit\'e Côte d’Azur, Nice, France}
\author[0000-0002-7352-7941]{Kevin B. Stevenson}
\affiliation{Johns Hopkins University Applied Physics Laboratory, 11100 Johns Hopkins Rd, Laurel, MD, 20723, USA}
\author[0000-0003-2415-2191]{Julien de Wit}
\affiliation{Department of Earth, Atmospheric and Planetary Sciences, Massachusetts Institute of Technology, Cambridge, MA 02139, USA}

\author[0009-0000-6113-0157]{Saugata Barat}
\affiliation{Anton Pannekoek Institute for Astronomy, University of Amsterdam,
Science Park 904, 1098 XH,
Amsterdam, the Netherlands}
\author{Jonathan Fortney}
\affiliation{Department of Astronomy and Astrophysics, 
University of California, 
Santa Cruz, 95064, CA, USA}
\author[0000-0003-3759-9080]{Tiffany Kataria}
\affiliation{NASA Jet Propulsion Laboratory, California Institute of Technology, Pasadena, CA 91109, USA}
\author{Michael Line}
\affiliation{School of Earth and Space Exploration, Arizona State University, Tempe, AZ 85287, USA}

\begin{abstract}
The extreme environments of transiting close-in exoplanets in highly-eccentric orbits are ideal for testing exo-climate physics. Spectroscopically resolved phase curves not only allow for the characterization of their thermal response to irradiation changes but also unveil phase-dependent atmospheric chemistry and dynamics.
We observed a partial phase curve of the highly-eccentric close-in giant planet HAT-P-2b ($e=0.51,M=9M_{\rm{Jup}}$) with the Wide Field Camera 3 aboard the \textit{Hubble Space Telescope}. Using these data, we updated the planet's orbital parameters and radius, and retrieved high-frequency pulsations consistent with the planet-induced pulsations reported in Spitzer data. We found that the peak in planetary flux occurred at $6.7\pm0.6$\,hr after periastron, with a heating and cooling timescales of $9.0^{+3.5}_{-2.1}$\,hr, and $3.6^{+0.7}_{-0.6}$\,hr, respectively. We compare the light-curve to various 1-dimensional and 3-dimensional forward models, varying the planet's chemical composition.
The strong contrast in flux increase and decrease timescales before and after periapse indicates an opacity term that emerges during the planet's heating phase, potentially due to more H$^{-}$ than expected from chemical equilibrium models.
The phase-resolved spectra are largely featureless, that we interpret as indicative an inhomogeneous dayside.
However, we identified an anomalously high flux in the spectroscopic bin coinciding with the hydrogen Paschen $\beta$ line and that is likely connected to the planet's orbit. We interpret this as due to shock heating of the upper atmosphere given the short timescale involved, or evidence for other star-planet interactions. 
       
\end{abstract}


\section{Introduction}
\label{sec:intro}
The extreme nature of close-in gas giants provides a unique window into the structure and dynamics of exo-atmospheres. Due to their short periods, relative brightness, and large atmospheric scale heights close-in giants provide excellent opportunities to improve our understanding of exo-atmospheres through the comparison of their observations to models. Close-in exoplanets receive intense irradiation from their host stars. Their atmosphere's ability to absorb, circulate and re-emit control the evolution of the planet and their atmosphere's thermal structure. 

Atmospheric composition, chemistry and dynamics are intrinsically connected. Together, they determine how a planet responds to the host star's irradiation \citep{Showman2009, Lewis2014}. These atmospheric properties set the radiative, advective and chemical timescales; three interconnected timescales that are key to understanding the structure of exo-atmospheres. They control how an extreme variation in energy gets absorbed/emitted by the atmosphere, how the energy is redistributed over the planet and how the atmospheric composition changes \citep[e.g.][]{Handbook2018}. These timescales are fundamental variables in dynamical atmospheric models \citep[e.g.][]{Iro2010, Visscher2012,  Tsai2023}. As such they are key to understanding the population of gaseous exoplanets, but they are \textit{a priori} unknown.

Recent studies by e.g. \citet{Ehrenreich2020, Mikal-Evans2022} are starting to unveil the intrinsic 3D nature of exoplanets by finding chemical gradients across the atmosphere. Advancements in high resolution spectroscopy techniques \citep[e.g.][]{Pino2022} and the unprecedented precision and wavelength coverage of JWST necessitate the development and testing of 3D atmospheric models \citep[e.g.][]{Caldas2019, Beltz2021,Wardenier2021, Challener2022}. 

Transiting close-in exoplanets in highly-eccentric orbits provide the ideal laboratory to test exo-atmospheric physics because their environments change rapdily through the extreme variations in instellation \citep[e.g.][]{deWit2016}. By spectrally resolving their phase curves one can not only characterize their response to the change in irradiation, but one can also reveal their phase-dependent chemistry and dynamics \citep{Lewis2014, Mayorga2021, Tsai2023}

A distinct atmospheric property that arises for Hot Jupiters hotter than $\sim$1700\,K is the presence of a thermal inversion \citep{Hubeny2003, Baxter2020}. An atmospheric layer heats up with respect to the surrounding layers when its optical opacities exceed its infrared opacities. This creates a region where the temperature increases with decreasing pressure. Likely agents for this type of atmospheric heating in Hot Jupiters are TiO, VO, and H$^{-}$ \citep{Hubeny2003, Fortney2008, Lothringer2018}. When the chemical timescale of the production of these molecules is short enough, eccentric planets that cross the $\sim$1700\,K local equilibrium temperature boundary may have a transient thermal inversion in their atmosphere \citep{Iro2010, Lewis2014}.

\begin{figure}
    \centering
    \includegraphics[width=\columnwidth]{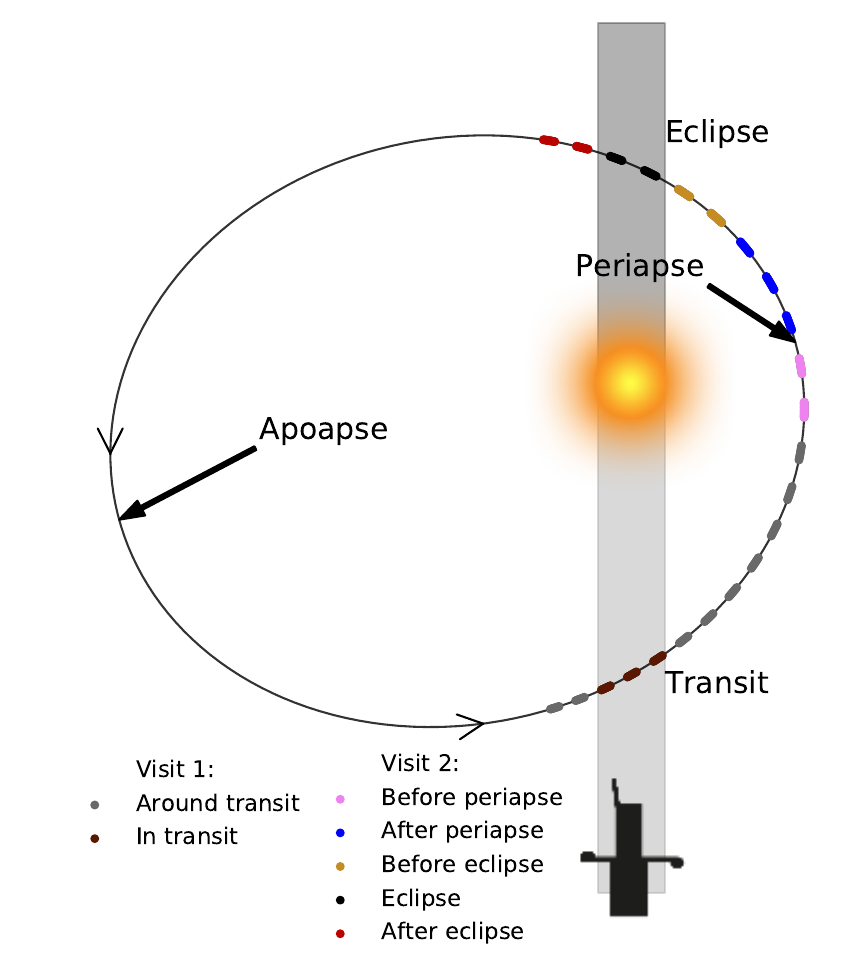}
    \caption{Visualization of HAT-P-2b's orbit, the locations of transit, eclipse and the phases that we observed. The planet, star and orbit are to scale. We color code the HST orbits to correspond to the colors in the subsequent figures. 
    The two visits were taken consecutively.
    }
    \label{fig:observations}
\end{figure}

HAT-P-2b is a $9.04\pm0.1$\,$M_{\rm{Jup}}$ planet in a highly-eccentric $(e=0.51023)$ orbit \citep{Bakos2007, Southworth2010,deWit2017, deBeurs2023}. Its short period (5.63 days, $a/R_s\approx9$) causes the planet to heat up from a typical Hot Jupiter ($\sim$1300\,K) to an Ultra-hot Jupiter ($\sim$2400\,K) and cool down again in two Earth days \citep{Lewis2013}. Its argument of periastron \citep[$\omega_{\star}\approx190$;] []{deWit2017, deBeurs2023}  allows us to observe primary transit just before the heating phase and the planet's secondary eclipse as it is cooling down quickly (see Figure \ref{fig:observations}).

\citet{Lewis2013} and \citet{deWit2017} observed HAT-P-2b's phase curve with the Spitzer Space telescope. At the Spitzer wavelengths (3.6, 4.5 and 8.0\,$\mu$m) they find that the planet's flux peaks 4-6\,hours after periastron.
This shows that the planet's atmosphere needs time to respond to the increase in incoming radiation. \citet{Lewis2013} therefore constrain the radiative timescale at mid-IR photospheric levels near periapse to be between 2 and 8 hours.

Additionally, \citet{deWit2017} found evidence for planet-induced stellar pulsations \citep{Bryan2024}. They found two modes with respective amplitudes of $35\pm7$\,ppm and $28\pm6$\,ppm and frequencies of $\sim$79 and $\sim$91 times the orbital frequency. Furthermore, \citet{deWit2017} found hints of apsidal precession and an evolving eccentricity in the radial velocity data. \citet{deBeurs2023} extended the baseline of radial velocity observations from 9 to 14 years and found that hints of this orbital evolution appear to persist, but additional follow-up observations are needed for confirmation. \citet{deBeurs2023} also confirmed the presence of a long-term companion of $10.7_{-2.2}^{+5.2}$\,$M_{\rm{Jup}}$ in an orbit of 23.3 years.

HAT-P-2b's phase curve is too long to feasibly observe in full. Partial phase curves of eccentric exoplanets have been observed before \citep{Lewis2013, deWit2016}, but never spectroscopically. \citet{Kreidberg2018} spectroscopically observed two partial phase curves of the non-eccentric WASP-103b. Full phase curves include a second secondary eclipse, which sets a baseline to which stellar and telescopic variability can be compared. Partial phase curves lack an accurate baseline. Planetary day-long variations in light-curve fits therefore become degenerate with instrumental effects and/or stellar variability \citep{Lewis2013, deWit2017}. To this end, \citet{Arcangeli2021} devised a new method to spectroscopically analyze partial phase curves that does not require \textit{a priori} knowledge of the instrumental systematics and therefore do not require a continuous observation of the planet.

The structure of this paper is as follows: in Section \ref{sec:obs} we first discuss the data used for this paper. We discuss how we convert the data into astrophysical signals using both light-curve fits as well as a common-mode method that retrieves phase-resolved spectra. In Section \ref{sec:atm_models} we shortly describe the atmospheric models to which we compare the data. Sections \ref{sec:LCresults} and \ref{sec:results:partialphase} present the results obtained from the light-curve fits and the phase-resolved spectra respectively.
We close off with a discussion on the astrophysical interpretation of our results in Section \ref{sec:discussion}.

\section{Data analysis}
\label{sec:obs}
\subsection{Observations and data reduction}
\label{sec:analysis_obs}
Our team observed HAT-P-2b for one primary transit, one secondary eclipse, and the partial phase curve in between for Program GO 16194 (PI: J.M. D\'esert, DOI:\dataset[10.17909/xnvc-as22]{https://doi.org/10.17909/xnvc-as22}) with WFC3 in December 2020. We visualize the timing of the observations and HAT-P-2b's orbit in Figure \ref{fig:observations}. Table \ref{tab:phases} lists the planetary orbital phases, orbital distances, and the time after mid-transit at which the observations were made, sorted into 7 groups. The equilibrium temperatures at apoapse is $1220$\,K at the furthest observed orbital distance $T_{\rm{eq}}=1540$\,K, and at periapse $T_{\rm{eq}}=2150$\,K. The data were obtained with the G141 grism, covering 1.1 to 1.7\,$\mu$m, using the bi-directional spatial scanning technique. We used the 512 $\times$ 512 subarray and the \texttt{SPARS25}, \texttt{NSAMP = 8} readout mode.

\begin{figure*}
    \centering
    \includegraphics[width=\textwidth]{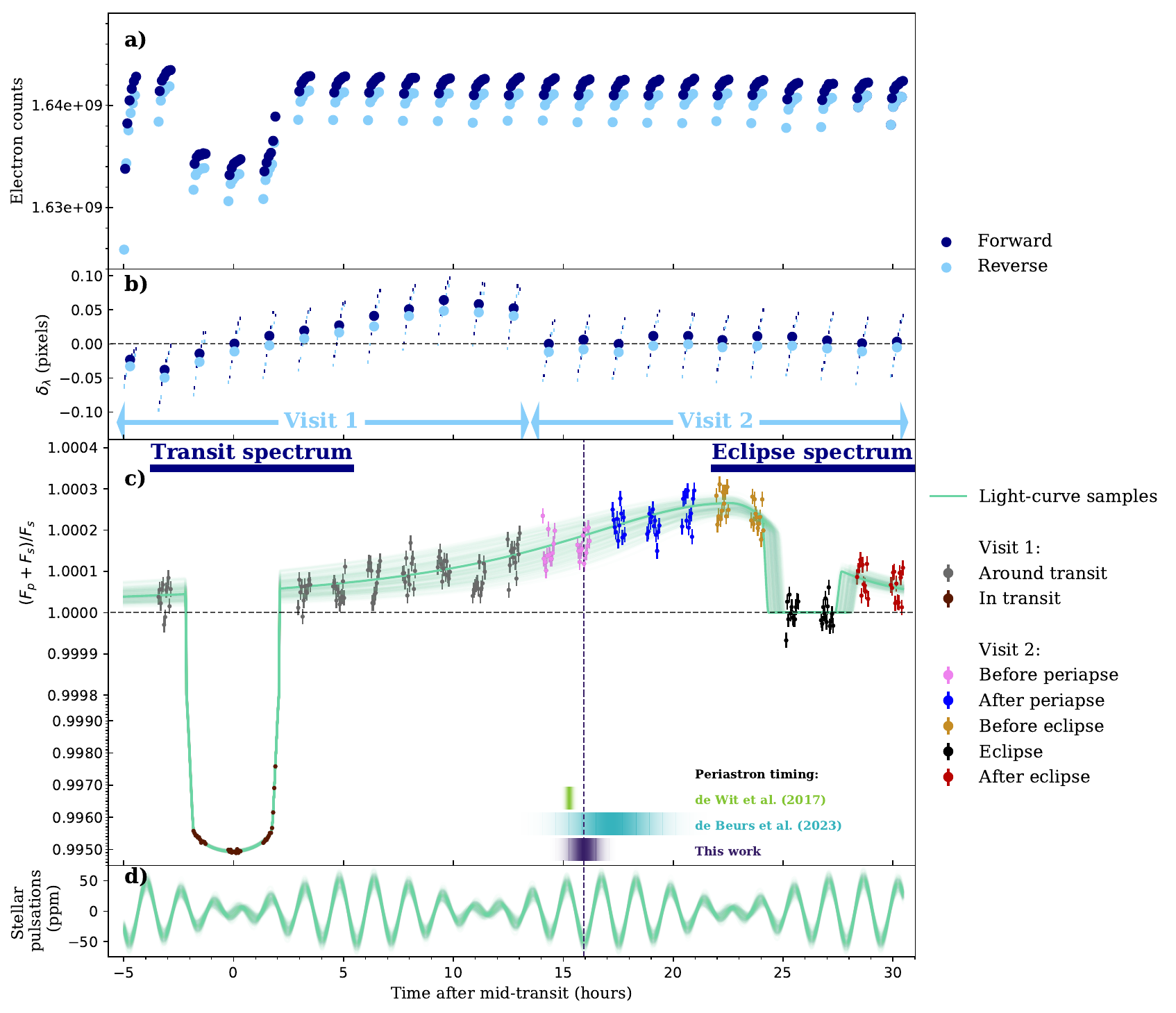}
    \caption{HAT-P-2b's WFC3 G141 light-curve starting just before transit and ending just after secondary eclipse. \\
    \textbf{a)} Raw WFC3 G141 light-curve separated by scan direction. The flux levels of the two scan directions are different due to the upstream-downstream effect \citep{McCullough2012}. Note that the first exposure of the penultimate orbit is missing: we omitted it because of a satellite crossing event.\\   
    \textbf{b)} Relative drift in pixels of each exposure on the WFC3 detector in the dispersion direction with respect to the reference exposure ($\delta_\lambda$). Large dots mark the average over $\delta_\lambda$ for each orbit.\\
    \textbf{c)} Telescope systematics-removed white light-curve data and best-fit planetary flux model ($\Phi(\mathbf{t}) + M(\mathbf{t})$) relative to the mean stellar flux. The first orbit has already been removed. The subsequent orbits have the same color coding as in Figure \ref{fig:observations}. We also show 200 posterior samples from the light-curve fits. We bolded the best-fit, which is pushed to an early eclipse by the priors. Note that the scale of the y-axis is different above and below $(F_p+F_s)/F_s=0.9998$. For the white light-curve data we removed the best-fit telescope systematics. These are correlated with the planetary- and stellar fit parameters. Different light-curve posterior samples correspond to different telescope systematics and would they therefore alter the representation of the data in this figure slightly.
    We highlight the orbits that we used for the transit and eclipse spectra. In the bottom of this subfigure we plot the posterior distributions for the timing of periastron from \citet{deWit2017}, \citet{deBeurs2023}, and this work. The median of the periastron posterior of this work is denoted with a vertical dashed line. \\
    \textbf{d)} Stellar light-curve $\Psi(\mathbf{t})$ showing stellar pulsations from the 200 posterior samples shown in subfigure c. The scaling of this subplot is equal to the scaling of the upper part of subfigure c. The retrieved pulsation phases resemble the pulsation phases found in Spitzer data 5--9 years prior to these observations \citep{deWit2017}.
    }
    \label{fig:rawdata}
\end{figure*}

\begin{table}[]
\centering
\begin{tabular}{lccccl}
Visit & HST & Phase & $r/R_s$  & Time after & Name \\
&orbit & & & mid-transit &\\
&number & & & (hrs) &\\
\hline
\hline
\multirow{12}{*}{1} & \cellcolor{mygrey!20} 1 & -0.035 & 8.7 & -4.7 & Around transit\\
\cline{2-6}
 & \cellcolor{mybrown!40} 2 & -0.023 & 8.3 & -3.1 & \multirow{3}{*}{In transit}   \\
 \cline{2-5}
 & \cellcolor{mybrown!40} 3 & -0.012& 7.9  & 1.5 &    \\
  \cline{2-5}
 & \cellcolor{mybrown!40} 4 & 0.001 & 7.6 & 0.1 &   \\
\cline{2-6}
 & \cellcolor{mygrey!20} 5 & 0.012 & 7.2 & 1.7 & \multirow{8}{*}{Around transit}  \\
  \cline{2-5}
 & \cellcolor{mygrey!20} 6 & 0.024 & 6.7 & 3.3 &   \\
  \cline{2-5}
 & \cellcolor{mygrey!20} 7 & 0.036 & 6.3 & 4.9 &   \\
  \cline{2-5}
 & \cellcolor{mygrey!20} 8 & 0.048 & 5.9 & 6.4 &   \\
  \cline{2-5}
 & \cellcolor{mygrey!20} 9 & 0.059 & 5.6 & 8.0 &   \\
  \cline{2-5}
 & \cellcolor{mygrey!20} 10 & 0.071 & 5.2 & 9.6 &   \\
  \cline{2-5}
 & \cellcolor{mygrey!20} 11 & 0.083 & 4.9 & 11.2 &   \\
  \cline{2-5}
 & \cellcolor{mygrey!20} 12 & 0.095 & 4.7 & 12.8 &   \\
   
\hline
\multirow{9}{*}{2} & \cellcolor{myviolet!40} 13 &  0.106 & 4.5 & 14.4    & \multirow{2}{*}{Before periapse}                                                                                                                                   \\
\cline{2-5}
 & \cellcolor{myviolet!40} 14 &  0.118 & 4.5 & 16.0    &                                                                                                                                                   \\
\cline{2-6}

 & \cellcolor{myblue!40} 15 &  0.130 & 4.5 & 17.6    & \multirow{3}{*}{After periapse}                                                                                                                                      \\
\cline{2-5}
 & \cellcolor{myblue!40} 16 &  0.142 & 4.7 & 19.1    &                                                                                                                                              \\
\cline{2-5}

 & \cellcolor{myblue!40} 17 &  0.153 & 4.9 & 20.7    &                                                                                                                                                 \\
\cline{2-6}
 & \cellcolor{mylime!40} 18 &  0.165 & 5.2 & 22.3    & \multirow{2}{*}{Before eclipse}                                                                                                                                     \\
\cline{2-5}
 & \cellcolor{mylime!40} 19 &  0.177 & 5.6 & 23.9    &                                                                                                                                                   \\
\cline{2-6}
 & \cellcolor{mygrey!80} 20 &  0.188 &  6.0 & 25.5  & \multirow{2}{*}{Eclipse}                                                                                                                                     \\
\cline{2-5}
 & \cellcolor{mygrey!80} 21 & 0.200 &  6.3 & 27.0   &      \\
\cline{2-6}
 & \cellcolor{myred!40} 22 &  0.212 & 6.8 & 28.6    & \multirow{2}{*}{After eclipse}                                                                                                                                     \\
\cline{2-5}
 & \cellcolor{myred!40} 23 & 0.224 & 7.2 & 30.2    &                                                                   \\
\hline

\end{tabular}

\caption{Average planetary orbital phase, orbital distance in stellar radii, and time after mid-transit at which each HST orbit was observed. We give each group of orbits a color and a name that we will use throughout this work. }
\label{tab:phases}
\end{table}

The first exposure of the penultimate orbit was affected by a satellite crossing event. Like the second satellite crossing in \citet{Fu2020} and the single satellite crossing in \citet{Jacobs2022}, we found no signs of its persistence impacting the remainder of the orbit. We therefore simply discarded the exposure with the satellite crossing. 

We used the data reduction pipeline developed by \citet{Arcangeli2018} and updated by \citet{Jacobs2022, Jacobs2023} to convert the raw data into spectra. We picked the sixth in-eclipse exposure as our reference exposure to which we compare each exposure's wavelength solution. 
Figure \ref{fig:rawdata}a shows the resultant raw light-curve averaged across all spectral channels in electrons per second. We call this the white light-curve.

The observations were split into two visits that were taken consecutively and without a gap, allowing for a gyro bias update in between the visits. This resulted in different telescope systematics in the flux measurements between the two visits. As the second visit has both the best pointing accuracy as well as the most interesting information, we split it further into five almost equally-sized groups of data visualized in Figure \ref{fig:observations}.

The South-Atlantic Anomaly (SAA) is a region above South America and the southern Atlantic Ocean where the cosmic ray flux levels are elevated compered to anywhere else on Earth due to a local dip in the Earth's magnetic field. These cosmic rays strongly contaminate data, strongly impairing their reliability. Therefore, generally, no observations are taken while crossing the SAA. HST crossed the SAA during Earth occultation starting with the eighth orbit until the fifteenth orbit of the observations presented in this work. Although HST did not take observations during the SAA crossing, the persistence of the elevated levels of cosmic rays may adversely affect the measured flux levels and the precision of the exposures taken after exiting the SAA in those orbits \citep{Barker2009, Stevenson2019}.

HST provides excellent pointing accuracy. But, for extended, multi-orbit observations, the telescope tends to drift ${\sim}5$\,mas\,${\approx}\,0.04$\,pixels during a 12-orbit visit \citep{Stevenson2019}. We cross-correlated the drop-off of the grism's sensitivity curve for each exposure to the position of the drop-off in the reference exposure to find $\delta_\lambda$, the band-integrated drift of the spectrum on the detector in the wavelength/dispersion direction. We show $\delta_\lambda$ in Figure \ref{fig:rawdata}b. There is a clear difference in alignment between the first visit and the second visit: the telescope drifted a total of ${\sim}0.1$ pixel when averaged per orbit during the first visit, but it remained remarkably stable during the second visit. The measured intra-orbit drift may be due to telescopic breathing, rather than a genuine drift of the telescope. The root mean squared of $\delta_\lambda$ is 0.05 for the first visit and 0.03 for the second visit, both well inside the bounds defined by \citet{Stevenson2019} to be considered a successful observation.

\subsection{White light-curve fitting}
\label{sec:WL}
Figure \ref{fig:rawdata}a shows that time-dependent systematics dominate the WFC3 light-curve. They are strongest in the first orbit, which we therefore discarded. We modeled the flux $F(\mathbf{t})$ of the remaining data with:

\begin{equation}
\label{eq:xi}
F(\mathbf{t}) = T(\mathbf{t}) \times  \Theta_{v, d}(\mathbf{t})  \times (\Phi(\mathbf{t})\, + M(\mathbf{t}) \times \Psi(\mathbf{t})\,)
\end{equation}
where $\mathbf{t}$ is a vector of observation times, $T(\mathbf{t})$ models the orbit-long telescope systematics, $\Theta_{v, d}(\mathbf{t})$ models the visit-long telescope systematics, $\Phi(\mathbf{t})$ models the planetary phase curve, $M(\mathbf{t})$ models the planetary primary transit using \texttt{batman} \citep{Kreidberg2015}, and $\Psi(\mathbf{t})$ models the stellar light-curve. The subscripts $v$ and $d$ denote whether a parameter is a function of telescope visit, and scan direction respectively.

Firstly, we modeled the telescope systematics, $T(\mathbf{t})$, with the physically-motivated coupled differential equations from the \texttt{RECTE} charge trap model \citep{Zhou2017}. This requires four fit parameters: the number of initially filled fast/slow charge traps ($E_{f/s,0}$) and the number of fast/slow charge traps that are filled in between orbits ($\Delta E_{f/s}$). These parameters are shared between scan directions and visits. Having different \texttt{RECTE} parameters for the first and second visit does not improve the fits significantly. 

Secondly, to model the visit-long telescope systematics we used a second-order polynomial for each visit separately:
\begin{equation}
\Theta_{v, d}(\mathbf{t}) = (f_{v, d} + V_{1_{v, d}}(\mathbf{t} - t_{f_v}) + V_{2_{v, d}}(\mathbf{t} - t_{f_v})^2) \times (1 + b\, \delta_\lambda)
\end{equation}
where $f_{v, d}$ is a normalization constant, $V_{1_{v, d}}$ is a visit-long linear slope, $V_{2_{v, d}}$ is a visit-long second-order polynomial term, $t_{f_v}$ is the time of the first exposure of the visit, $b$ is a fit parameter and $\delta_\lambda$ is the band-integrated drift of the spectrum on the detector in the wavelength/dispersion direction seen in Figure \ref{fig:rawdata}b. 
We tested a model where we set $V_{2_{v, d}}=0$ for either of the two visits, but this was disfavored by the Bayesian information criterion (BIC) for the white light-curve. Such model was disfavored for the first visit by $\Delta$(BIC)$=314$ and it was disfavored for the second visit by $\Delta$(BIC)$=39$. A model with a single unified second-order polynomial for both visits was disfavored by $\Delta$(BIC)$=99$.  
A $\Delta$(BIC)$>10$ is generally considered as very strong evidence and corresponds to a model preference of approximately 150:1 \citep{KassRaftery1995}.
This is a clear indication that the telescope systematics were different between the two visits.

\citet{Haynes2015} improved their WFC3 light-curve fits by correcting for the spectral drift in the wavelength direction, $\delta_\lambda$. They did so by either shifting each exposure's spectrum with respect to a reference spectrum during the data reduction phase of the analysis, or by decorrelating against $\delta_\lambda$ by adding an extra term $(1 + b\, \delta_\lambda)$ to their light-curve model \citep[see also: ][]{Wakeford2016, Tsiaras2016a, Barat2023}. We chose the first method and performed it during our data reduction phase. However, despite that, we found that by additionally adding the extra $b$-parameter the BIC value of the full white light-curve fit improved by 71. A fit parameter to account for the shift of the spectrum in the spatial scan direction did not improve the fits. We note from Figure \ref{fig:rawdata}b that the repetitive intra-orbit shape of $\delta_\lambda$ resembles the hook-like pattern in Figure \ref{fig:rawdata}a. Also, a corner-corner plot of the posteriors (Figure \ref{fig:cornercorner} in Appendix \ref{app:additional_figs}) shows clear correlations between $b$ and the \texttt{RECTE} parameters. Therefore, the fit parameter $b$ may, in fact, be tracing telescope breathing/charge trap systematics left uncorrected for by \texttt{RECTE} rather than a true dependence on the drift of the spectrum. Similarly, $V_{1_{v, d}}$ and $V_{2_{v, d}}$ may also be tracing a change in stellar brightness on a timescale longer than the duration of the telescope visits.

Thirdly, to model HAT-P-2b's phase curve we followed \citet{deWit2017} and used an asymmetric Lorentzian model:
\begin{eqnarray}
\label{eq:Phi}
\Phi(\mathbf{t}) &=& 
\begin{cases} 
    0 & \text{if: in eclipse} \\
    F_p/F_{s, \rm{min}}+\frac{c_1}{u(\mathbf{t})^2+1} & \text{if: out eclipse} 
\end{cases}
\\
u(\mathbf{t})
&=&
\begin{cases} 
    (\mathbf{t} - t_{\textrm{per}} - c_2)/c_3 & \text{if: $\mathbf{t}- t_{\textrm{per}}<c_2$;}  \\ 
    (\mathbf{t}- t_{\textrm{per}} - c_2)/c_4 & \text{if: $\mathbf{t}- t_{\textrm{per}}>c_2$} 
\end{cases}
\end{eqnarray}
where $F_p/F_{s, \rm{min}}$ is the minimum flux of the phase curve, $t_{\textrm{per}}$ is the time of periastron passage and $c_1$, $c_2$, $c_3$, and $c_4$ are free parameters. $c_1$ is the maximum flux of the phase curve minus the minimum flux; $c_2$ is the difference in time between periastron and the time of maximum flux; $c_3$ is the flux increase timescale; and $c_4$ is the flux decrease timescale. The timing of secondary eclipse and periastron passage are calculated from the planet's orbital parameters with \texttt{batman}. We investigated using a phase curve model based on harmonics in orbital phase \citep{CowanAgol2008}, like used by \citet{Lewis2013}. Statistically, it provides a better fit than an asymmetric Lorentzian, but it creates a double humped light curve with its brightest peak during transit. We discard this harmonics-based fit as improbable.

Fourthly, the transit light-curve, $M(\mathbf{t})$, requires the input of HAT-P-2b's orbital parameters. \citet{deBeurs2023} found hints that the eccentricity and argument or periastron may be significantly evolving, which matches what would be expected from stellar evolution models that incorporate tidal planet-star interactions \citep{Bryan2024}. We therefore plugged six fit parameters into \texttt{batman}: the planet's radius in units of stellar radii, $R_{p}/R_s$; the mid-transit time, $t_0$; the planet's orbital inclination, $i$; the orbital eccentricity, $e$; its argument of periastron, $\omega_{\star}$; the orbital semi-major axis in units of stellar radii, $a/R_s$; and the linear limb-darkening coefficient $u$.

Lastly, to model the stellar light-curve, $\Psi(\mathbf{t})$, we used the two stellar pulsation modes discovered by \citet{deWit2017}:
\begin{equation}
    \Psi(\mathbf{t}) = 1 + A_1 \cos\left( \frac{2\pi}{P_1} (\mathbf{t} - t_0 + \phi_1)\right) + A_2 \cos\left( \frac{2\pi}{P_2} (\mathbf{t} - t_0 + \phi_2)\right)
\end{equation}
where $A_{1,2}$ are the pulsation amplitudes, $P_{1,2}$ the pulsation periods, and $\phi_{1,2}$ the pulsation phases. The WFC3 observations were made years after the Spitzer observations, rendering the priors from \citet{deWit2017} uninformative for these WFC3 observations. We also fitted for $A_{1,2}$ and $\phi_{1,2}$, but we fixed $P_{1,2}$ to the fractions of the planet's orbital period, $P_{\rm{orb}}$, found by \citet{deWit2017}: $P_1=\frac{1}{79}P_{\rm{orb}}=103$\,min, and $P_2=\frac{1}{91}P_{\rm{orb}}=89$\,min. Coincidentally, the two pulsation periods found by \citet{deWit2017} are extremely close to the orbital period of HST: ${\sim}96$\,min. This could cause some degeneracies between the pulsations and the orbit-long telescope systematics. Due to this, and because of the limited observed phase range, we did not search for other pulsation periods or modes. We gain confidence of including the two pulsation modes at these exact periods through the large difference in BIC that including the pulsations make: ${\Delta}\rm{BIC}=21$. 

Without pulsations the transit depth would be $1.2\sigma$ deeper, the flux increase timescale, eclipse depths would be $2\sigma$ shallower, $c_3$, would be $1.8\sigma$ shorter, and the flux decrease timescale, $c_4$, would be $1.1\sigma$ longer.

\subsection{Transit and eclipse spectra}
\label{sec:analysis:t_and_e}
To obtain transmission and eclipse spectra we split the G141 spectrum into 17 spectrophotometric bins that are 7-pixels wide and we fitted the spectral light-curves separately. We made sure that the most significant stellar line (hydrogen Paschen $\beta$ at 1.282\,$\mu$m) is fully encompassed by one bin and is therefore not split over multiple bins. This helps to decrease the impact of the drifting of the spectrum on the detector.

Because of the effects of the SAA and the enlarged uncertainties associated with fitting the full partial phase curve, we measured the transit and eclipse spectra by focusing on a subset of the observations: we used the one HST orbit preceding transit, the three in-transit orbits, and the two orbits following transit to determine the transit spectrum. For the eclipse spectrum we used the two orbits before eclipse, the two orbits in eclipse and the two orbits after eclipse. This is illustrated in Figure \ref{fig:rawdata}c. 

We fitted the spectral light-curves with a similar model as the white light-curve fit described in the preceding section, but we fixed a number of parameters to the values found for the white light-curve: the orbital parameters, the spectral drift fit parameter $b$, the planetary light-curve parameters $F_p/F_{s, \rm{min}}$, $c_2$, $c_3$, and $c_4$, and the stellar pulsation parameters. This means that the only fit parameter left for the planetary light-curve (Equation \ref{eq:Phi}) is the maximum flux, $c_1$. 
We therefore kept $F_p/F_{s, \rm{min}}$ at the value retrieved for the white light-curve. We obtained the secondary eclipse depth by evaluating Equation \ref{eq:Phi} at the time of secondary eclipse, $t_{\rm{sec}}$. Because of the shorter baseline, we reduced the second-order polynomial visit-long baseline for the second visit to a linear baseline. The first visit still requires a second-order polynomial to properly describe the visit-long telescope systematics. 

Despite the light-curve no longer being continuous, we kept $E_{f/s,0}$, $\Delta E_{f/s}$, and $b$ shared between the transit and the eclipse data because we did not find any significant difference to these parameters between the visits. Sharing the telescope systematics parameters allows us to fit both data sets simultaneously to decrease the uncertainties on the telescope systematics and therefore increase the signal-to-noise ratio on the transit and eclipse depths. If we solely fit the secondary eclipse, the resulting spectrum differs from the combined fit by on average $0.5\sigma$ while the error bars are larger by 3\%.

\begin{figure}
    \centering
    \includegraphics[width=\columnwidth]{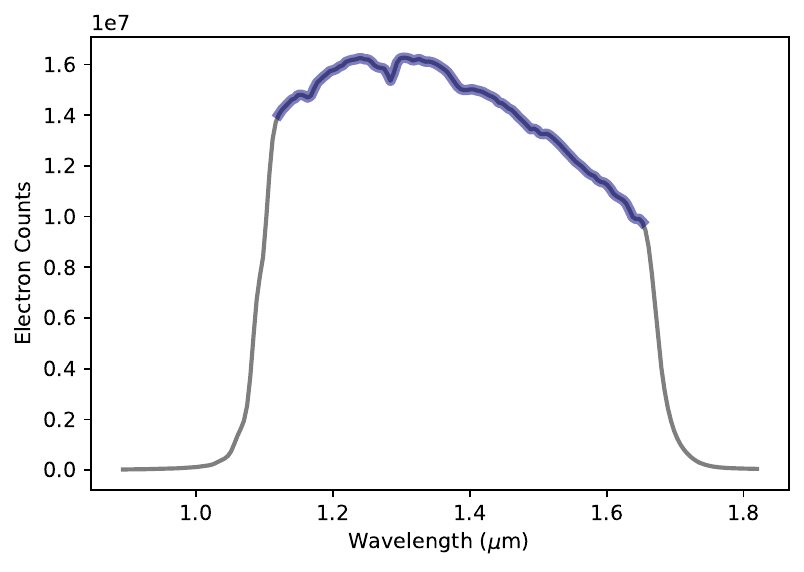}
    \caption{Average observed spectral stellar electron count rate per exposure during the occultation of HAT-P-2b. It is a convolution of the actual stellar spectrum and the G141 transmission function. The wavelengths that are used for this work are highlighted in blue. The observed count rate at the longest wavelength is only 60\% of the peak count rate, requiring wavelength-dependent telescope systematics parameters. 
    }
    \label{fig:stelflux}
\end{figure}

The stellar flux varies significantly between wavelengths, see Figure \ref{fig:stelflux}. Therefore, a different number of charge traps are filled at each wavelength, and the \texttt{RECTE} parameters are therefore required to be wavelength-dependent.
In Table \ref{tab:priors} we summarize the 25 fit parameters, the respective priors that were used for the white light-curve fit and their references (see Section \ref{sec:MCMCs}). We also denote which eleven parameters are wavelength-dependent and therefore used for the 17 spectral fits. Because the divergence in measured amplitude between the stellar pulsations 1.1-1.7\,$\mu$m in this work and $4.5$\,$\mu$m in \citet{deWit2017} is only modest, we assumed the pulsations to be monochromatic over the WFC3 G141 wavelength range.

\subsection{Uncertainties and priors}
\label{sec:MCMCs}
In order to estimate the errors on the fitted parameters and identify the degeneracies in the model we used a Markov Chain Monte Carlo (MCMC) approach using the open-source \texttt{emcee} code \citep{emcee}. For the white light-curve fits, the chains have 30,000 steps with 250 walkers and a burn-in phase of 5000 steps. For the spectral fits, each chain has 10,000 steps with 110 walkers and a burn-in phase of 1000 steps. We test for convergence by averaging auto-correlation time over all walkers \citep{GoodmanWeare2010} and ensure that this is 50 times smaller than the number of samples.

Running an MCMC is computationally expensive, so minimizing the number of fit parameters is key. We therefore first ran a quick Levenberg-Marquardt fit \citep[LM-fit;][]{Newville2014} to get a good estimate of the best-fit parameters. We assumed that the LM-fit accurately determined the correct ratio between the visit-long telescope systematics of the two scan directions $V_{1/2_{v,d}}$, and we fixed that ratio. In the MCMC fit we then fitted for $\Delta V_{1/2_{v}}$, the difference with the value found by the LM-fit for the Forward scan direction, simplifying the light-curve fits. Furthermore, we also eliminated $f_{v,d}$ from the MCMC fits. $f_{v,d}$ provides an average electron level flux to \texttt{RECTE}. However, a relative difference in electron counts of less than a percent has an insignificant effect on the orbit-long ramp shape. We therefore deemed the value of $f_{v,d}$ found by LM-fit good enough and simply renormalize the model to the data in the MCMC fits. 

Literature-provided values serve as priors in the MCMCs for some parameters. \citet{deWit2017} found the minimum Earth-facing, hemisphere-averaged temperature to be $1327\pm106$\,K, which corresponds to $F_p/F_{s, \rm{min}}=9\pm5$\,ppm at WFC3 G141 wavelengths. We doubled the uncertainty and used this as a prior on $F_p/F_{s, \rm{min}}$, but we disallowed negative planetary fluxes.

\begin{deluxetable*}{ll|l||l|l|c|l}
\tablewidth{0pc}
\setlength{\tabcolsep}{3pt}
\tabletypesize{\small}
\tablecaption{Sample priors and posteriors for the WFC3 light-curve fits of HAT-P-2b. \label{tab:priors} }
\tablehead{
\multicolumn{2}{c|}{Parameter}                          & Unit & Prior\tablenotemark{\scriptsize a}                         & Posterior                 & Included in & Reference                                   \\
                    &   &             &                               &                           & spectral fits? & for prior}
\startdata
 &&&&&& \\
\textbf{\textit{Occultation depths}} &&&&&& \\
\hline
Transit depth & $R_p^2/R_s^2$ & ppm                & $\mathcal{U}(3500, 5500)$     & $4873\pm 9$               & \cmark                     &                                                    \\
Eclipse depth & $F_p/F_{s,sec}$ & ppm                & \tablenotemark{\scriptsize b}     & $141^{+14}_{-15}$               &                     &                                                    \\
 &&&&&& \\
\textbf{\textit{Orbital parameters}} &&&&&& \\
\hline
Mid-transit time offset to [1] & $\Delta t_0$ & min               & $\mathcal{N}(0, 0.7469)$       & $2.44\pm0.19$    &                            & [1]              \\
Orbital inclination & $i$ & deg                          & $\mathcal{N}(86.16,0.26)$     & $86.20\pm0.13$      &                            & [2]                 \\
Orbital eccentricity & $e$ &                               & $\mathcal{N}(0.50381,0.0325)$ & $0.509\pm0.005$ &                            & [3]               \\
Argument of periastron & $\omega_{\star}$ & deg                     & $\mathcal{N}(194.7, 4.0)$     & $191.6^{+1.9}_{-2.0}$                 &                            & [3]               \\
Scaled semi-major axis & $a/R_s$ &                           & $\mathcal{N}(9.04,0.19)$      & $9.16\pm0.15$               &                            & [4]                 \\
Linear limb darkening parameter & $u$     &                           & $\mathcal{N}(0.297,0.011)$    & $0.228\pm0.004$           & \cmark                     & [5] \\
Time between mid-transit and periastron & $t_{per}$ & hr                & \tablenotemark{\scriptsize b}     & $15.9\pm0.5$              &                     &                                                    \\
 &&&&&& \\
\textbf{\textit{Planetary phase curve parameters}} &&&&&& \\
\hline
Minimum flux & $F_p/F_{s, \rm{min}}$ & ppm        & $\mathcal{N}(9, 10)$         & $13^{+10}_{-8}$          &                            &                                                    \\
Flux difference ($F_p/F_{s,\rm{max}} - F_p/F_{s,\rm{min}}$) & $c_1$  & ppm                        & $\mathcal{U}(0,10^5)$         & $245\pm17$         & \cmark                     &                                                    \\
Peak flux delay from periastron & $c_2$ & hr                         & $\mathcal{U}(0, 24)$          & $6.7\pm0.6$        &                            &                                                    \\
Flux increase timescale & $c_3$ & hr                        & $\mathcal{U}(0, 14.4)$        & $9.1^{+3.4}_{-2.1}$        &                            &                                                    \\
Flux decrease timescale & $c_4$ & hr                         & $\mathcal{U}(0, 24)$          & $3.9^{+0.7}_{-0.6}$               &                            &                                                    \\
 &&&&&& \\
\textbf{\textit{Telescope systematics}} &&&&&& \\
\hline
Initially filled slow charge traps & $E_{0,s}$ &  count                       & $\mathcal{U}(0, 1525.38)$     & $1261\pm5$                & \cmark                     & [6]                  \\
Initially filled fast charge traps & $E_{0,f}$  & count                       & $\mathcal{U}(0, 162.38)$      & $89.8\pm1.3$              & \cmark                     & [6]                  \\
Extra filled slow charge traps in between orbits & $\Delta E_s$ & count                     & $\mathcal{U}(0, 1525.38)$     & $142.8\pm2.6$                 & \cmark                     & [6]                  \\
Extra filled fast charge traps in between orbits &$\Delta E_f$ & count                     & $\mathcal{U}(0, 162.38)$      & $91.1\pm0.4$              & \cmark                     & [6]              \\
Visit-long first polynomial term visit 1 & $\Delta V_{1_1}$\tablenotemark{\scriptsize c} & ppm\,day$^{-1}$ & $\mathcal{U}(-3700,3700)$     & $-100\pm120$               & \cmark                     &                                                    \\
Visit-long first polynomial term visit 2 &$\Delta V_{1_2}$\tablenotemark{\scriptsize c} & ppm\,day$^{-1}$ & $\mathcal{U}(-10000, 10000)$  & $-60^{+130}_{-110}$                & \cmark                     &                                                    \\
Visit-long second polynomial term visit 1 &$\Delta V_{2_1}$\tablenotemark{\scriptsize c} & ppm\,day$^{-2}$ & $\mathcal{U}(-8000,8000)$     & $-140\pm120$               & \cmark                     &                                                    \\
Visit-long second polynomial term visit 2 &$\Delta V_{2_2}$\tablenotemark{\scriptsize c} & ppm\,day$^{-2}$ & $\mathcal{U}(-2500,2500)$     & $1070\pm130$        &                            &                                                    \\
Linear flux dependence on $\delta_\lambda$ & $b$ & ppm\,pixel$^{-1}$            & $\mathcal{U}(-10^{6},10^6)$  & $1400\pm160$              & \cmark                     &                                                    \\
 &&&&&& \\
\textbf{\textit{Stellar pulsation parameters}} &&&&&& \\
\hline
Amplitude pulsation mode 1 & $A_1$ & ppm                        & $\mathcal{U}(0,105)$          & $23\pm5$                  &                            &                                                    \\
Amplitude pulsation mode 2  & $A_2$ & ppm                        & $\mathcal{U}(0,84)$           & $29\pm5$                  &                            &                                                    \\
Phase pulsation mode 1\tablenotemark{\scriptsize d}   & $\phi_1$ & hr                      & $\mathcal{U}(0,1.71)$         & $0.99\pm0.05$             &                            &                                                    \\
Phase pulsation mode 2\tablenotemark{\scriptsize d}   & $\phi_2$ &  hr                       & $\mathcal{U}(0,1.49)$         & $1.15\pm0.04$             &                            & \\
\enddata
\textbf{Notes.}
\vspace{-0.2cm}\tablenotetext{\textrm{a}}{ The symbol $\mathcal{U}$ is used to denote uniform priors and $\mathcal{N}$ is used for Gaussian priors.}
\vspace{-0.2cm}\tablenotetext{\textrm{b}}{ This parameter was not fit directly, but calculated from the other fit parameters. It is only shown here for reference}
\vspace{-0.2cm}\tablenotetext{\textrm{c}}{ We fit for the difference with a second-order polynomial that was fit to the data prior to a run with an MCMC (see Section \ref{sec:MCMCs}).}
\vspace{-0.2cm}\tablenotetext{\textrm{d}}{ The pulsation phases were measured with respect to mid-transit.}
\tablenotetext{}{References:\\
{[1]}: \citet{Ivshina2022} \\
{[2]}: \citet{deWit2017}\\
{[3]}: \citet{deBeurs2023}\\
{[4]}: \citet{Patel2022}\\
{[5]}: \citet{Castelli2003, Bourque2021}\\
{[6]}: \citet{Zhou2017}}
\end{deluxetable*}
The average final precision on the spectroscopic transit depths and eclipse depths are 49\,ppm and 23\,ppm respectively. 
This large difference in precision is largely caused by the requirement of a quadratic trend to model the telescope systematics in the first visit compared to a linear trend for the second visit ($V_{2_{2,d}}=0$). 
A quadratic baseline has larger degeneracies with the transit depth. Also, the out-of-transit baseline is shorter than the out-of-eclipse baseline. Additionally, the \texttt{RECTE} parameters with regards to slowly trapped charges are mainly dependent on the pre- and in-transit orbits, adding extra degeneracies with the transit depth, but not to the eclipse depth.
We reach an average precision of 6\% and 8\% above photon noise for transit and eclipse respectively.

In the light-curve fits we used four out-of-eclipse orbits. If we had used only two out-of-eclipse orbits and only fit for the eclipse light curve, the average precision on the eclipse spectrum would have decreased from 23 to 33 ppm. So, by doubling the out-of-eclipse baseline and observing a 6-orbit eclipse, we decreased the uncertainties by ${\sim}\sqrt{2}$, which is equivalent to observing a 4-orbit eclipse twice.

\subsection{Partial phase curve}
\label{sec:analysis:phase}
The most commonly used technique to remove telescope systematics from phase curves is to use a light-curve fitting method similar to the ones used in the previous sections \citep[e.g.][]{Stevenson2014, Kreidberg2018}. Alternatively, the common-mode-based method developed by \citet{Arcangeli2021} allows us to extract the planetary spectrum at any orbital phase without fitting a light-curve. It would even allow us to extract the planetary spectrum if the light-curve were not continuous. 

We give a short summary of the common-mode based method: one divides the spectral light-curve by the white light-curve. Subsequently one takes the in-eclipse spectrum as the stellar spectrum and divides all other spectra by this. We then grouped the data by HST orbits. This way, one can retrieve the planetary spectrum up to a monochromatic constant if the following restrictions are satisfied:

Firstly, the host star's spectrum should be invariable. The stellar pulsations found by \citet{deWit2017} at $4.5$\,$\mu$m also seem to be present in these WFC3 data. However, we detected them to be largely of the same amplitude. 
The maximum difference in peak-to-trough pulsation amplitude between this work at $1.4$\,$\mu$m and \citet{deWit2017} at $4.5$\,$\mu$m is 120\,ppm at 3$\sigma$. The maximum color difference between the pulsation peaks and troughs is therefore ${\sim}40$\,ppm\,$\mu$m$^{-1}$. 
We injected this spectral slope 
into our derived spectra and found that it alters the fitted temperature by 100\,K, 
which is similar to the $1\sigma$ uncertainties on the fits themselves.

Secondly, the drift in pointing of the telescope must be sufficiently small, as this can alter the observed spectrum significantly. Figure \ref{fig:rawdata}b shows that the orbit-to-orbit variations of the drift of the spectrum are at most 0.1 pixels and 0.012 pixels for the first and second visit respectively. The spectral drift for the first visit is similar to the spectral drift for the transit visit in \citet{Arcangeli2021} which they found to be ill-suited for their common-mode method due to this spectral drift. The effect of the spectral drift for the second visit, however, is smaller than the spectral noise according to \citet{Arcangeli2021}.

Thirdly, the long-term telescope systematics must be monochromatic. HST's visit-long systematics are found to be wavelength-dependent \citep[see e.g.][]{Kreidberg2014, Stevenson2014, HuberFeely2022}. The telescope drift contributes to these systematics, but may not account for all. We tested this for the observations presented in this work by analyzing the wavelength-dependence of the visit-long polynomial coefficients, $V_{1_{v, d}}$ and $V_{2_{d}}$, in the transit and eclipse fits. These coefficients may account for both telescope systematics as well as star color variations. So, if these coefficients are wavelength independent, the first and third requirements for the common-mode method are simultaneously satisfied. 
However, we found that the single linear coefficient for the eclipse data, $V_{1_2}$ is wavelength dependent by $2.6\sigma$ at 650\,ppm\,day$^{-1}$\,$\mu$m$^{-1}$.  This amounts to a possible 1800\,K underestimate of a planetary blackbody temperature during the first orbit of the second visit. The first order polynomial term of the transit light curve is also wavelength dependent by $2.7\sigma$.
The caveat is that we analyzed the wavelength dependence of the visit-long trends using the transit and eclipse fits, which were only fit to a limited set of data (see Section \ref{sec:analysis:t_and_e}). For the full white light-curve fits an extra polynomial order is required to accurately describe the second visit baseline. Also, the wavelength dependence of $V_{1_2}$ may rather hint at inaccuracies in the modelling of the planetary phase curve in the transit-and-eclipse-only fits.

We therefore deem this common-mode method possibly applicable to the second visit, and unfit for the first visit because it violates the second and the third requirements.

In Section \ref{sec:analysis_obs} we noted that we excluded the first exposure of the penultimate orbit because of the satellite crossing event. For the common-mode spectral reduction of this orbit, we therefore also omitted the first exposure of each in-eclipse orbit, such that it is not biased by the wavelength dependence of HST's orbit-long systematics. For all other orbits we continued to use all exposures.

\section{Atmospheric models}
\label{sec:atm_models}
\subsection{General Circulation models}
\label{sec:GCM_models}
We compare our observed WFC3 G141 white-light partial phase curve and phase-resolved spectra for HAT-P-2b with predictions from general circulation models (GCMs) presented in \citet{Lewis2014}. These HAT-P-2b GCMs use the Substellar and Planetary Atmospheric Radiation and Circulation (SPARC) model \citep[][]{Showman2009} that employs the MITgcm \citep{Adcroft2004} to treat the atmospheric dynamics and the non-gray radiative transfer model of \citet{Marley1999}, which has been applied in a wide variety of exoplanet atmospheric studies \citep[e.g][]{Fortney2005,Mayorga2019}. The HAT-P-2b GCMs consider atmospheric compositions of one times (1$\times$) and five times (5$\times$) solar metallicity both with and without TiO and VO. Chemical abundances are computed throughout the GCM grid assuming instantaneous thermochemical equilibrium given local temperature and pressure conditions \citep{Lodders2002, Lodders2003}. An intrinsic effective temperature ($T_{\rm{int}}$) of 300~K and a pseudo-synchronous rotation period of 1.9462563~days is assumed for HAT-P-2b \citep{Lewis2014}.

Spectra as a function of orbital phase are produced from GCM outputs (temperature as a function of pressure along each column of the grid) that are recorded at regular intervals during the simulation. We track both the longitude of the substellar point and the longitude of an Earth observer throughout the simulation given the properties of HAT-P-2b's orbit \citep{Kataria2013, Lewis2014}. Emergent spectra from each of the GCM grid points on the hemisphere facing the Earth observer are calculated based on the local thermochemical structure and then averaged according to the methods presented in \citet{Fortney2006} to produce a single planetary emission spectrum as a function of orbital phase. These emission spectra are then convolved with the relevant instrument throughput function(s) to produce light curves and spectra as a function of planetary orbital phase and instrument bandpass.

\subsection{1D time-stepping models}
\label{sec:1D_time_step}
We used \texttt{EGP+} \citep{Mayorga2021} to model both clear and cloudy atmospheres. \texttt{EGP+} is a one-dimensional, time  stepping, atmospheric model for planets on eccentric orbits. This model was built on the framework of the radiative-convective equilibrium models of \citet{Marley1999, Ackerman2001} and the time stepping evolution of \citet{Robinson2014}. Each model is initiated from apoastron and run for approximately 10 orbits to ensure convergence to a quasi-periodic state under the assumptions of solar metallicity and C/O ratio. These models were run using the system parameters from \citet{Pal2010} and when applicable include MgSiO$_3$, MnS, Na$_2$S as condensates. We also run versions where we forcefully excluded TiO and VO, the primary agents for thermal inversions.

\subsection{1D free-chemistry models}
\label{sec:1D_free-chemistry_models}

We used the open-source atmospheric retrieval package POSEIDON \citep{MacDonald2017, MacDonald2023, Coulombe2023} to infer the temperature structure and chemical composition of the atmosphere at each orbital phase. For this analysis, we used a temperature-pressure (TP) profile designed for brown dwarfs \citep{Piette2020}, but modified it to allow for negative temperature gradients (thermal inversions). The free parameters of this temperature model are the atmospheric temperature at 1 bar pressure and temperature gradients between the following pressure nodes: $10^{-5}$, $10^{-3}$, $10^{-1}$, $10^1$, and $10^2$ bar. The priors on the temperature differences between the parameterized layers are -1000 – 1000 K except between the deepest layers, which is uniform between 0 – 1000 K as we expect the deep atmosphere to be non-inverted. The prior on the temperature at 1 bar is uniform between 400 -- 4000 K. We include the following opacity sources: H$_2$-H$_2$ and H$_2$-He collision-induced absorption \citep{Karman2019}, H$_2$O \citep{Polyansky2018}, H$^-$ \citep{John1988}, FeH \citep{Wende2010}, TiO \citep{McKemmish2019}, and VO \citep{McKemmish2016}. We fit for constant-with-pressure volume mixing ratios in log space for each of the chemical species, with uniform priors between -12 -- -1, and assume the remainder of the atmosphere is H$_2$ and He at the solar ratio. Finally, we fit for the planet's radius at $10^{-2}$ bar and a constant offset on the spectra (phase-curve spectra only), both constrained by Gaussian priors from the white-light analysis.

We calculate emission spectra using opacity sampling at a resolution of $R = 10,000$. Our model has 100 pressure layers from $10^{-5} - 10^2$ bar. For the stellar spectrum, we use a PHOENIX model \citep{Husser2013} assuming a stellar radius of 1.64 $R_\odot$, an effective surface temperature of 6414 K, a metallicity of 0.04, and a surface gravity of $\log(g) = 4.18$ \citep{Pal2010, Tsantaki2014} interpolated with pyMSG \citep{Townsend2023}. We explore the parameter space with \texttt{PyMultiNest}, a Python extension of \texttt{MultiNest} \citep{Feroz2009, Buchner2014} using 1,000 live points.

\section{Results on the light-curve fits}
\label{sec:LCresults}
We display the best-fit white light-curve as well as 200 posterior samples for the planetary light-curve in Figure \ref{fig:rawdata}c and for the stellar light-curve in Figure \ref{fig:rawdata}d. In Table \ref{tab:priors} we list these posteriors for all the fit parameters.

Notable in the white light-curve data of Figure \ref{fig:rawdata}c is that the flux of the penultimate orbit of the first visit seems to be anomalously low compared to the best-fit light-curve. Before the data of this orbit were taken, the telescope experienced the strongest SAA crossing of the entire observation. The penultimate orbit of the first visit is therefore likely most impacted by the negative effects from the SAA described in Section \ref{sec:analysis_obs}. We tested omitting this orbit from the fits, but that does not change the posteriors of any fit parameters at more than $1\sigma$. 
We note that the SAA may have affected more orbits than just the penultimate orbit of the first visit, albeit less visibly.

\subsection{Orbital parameters}
\label{sec:results:orbitalparams}
\begin{figure}
    \centering
    \includegraphics[width=\columnwidth]{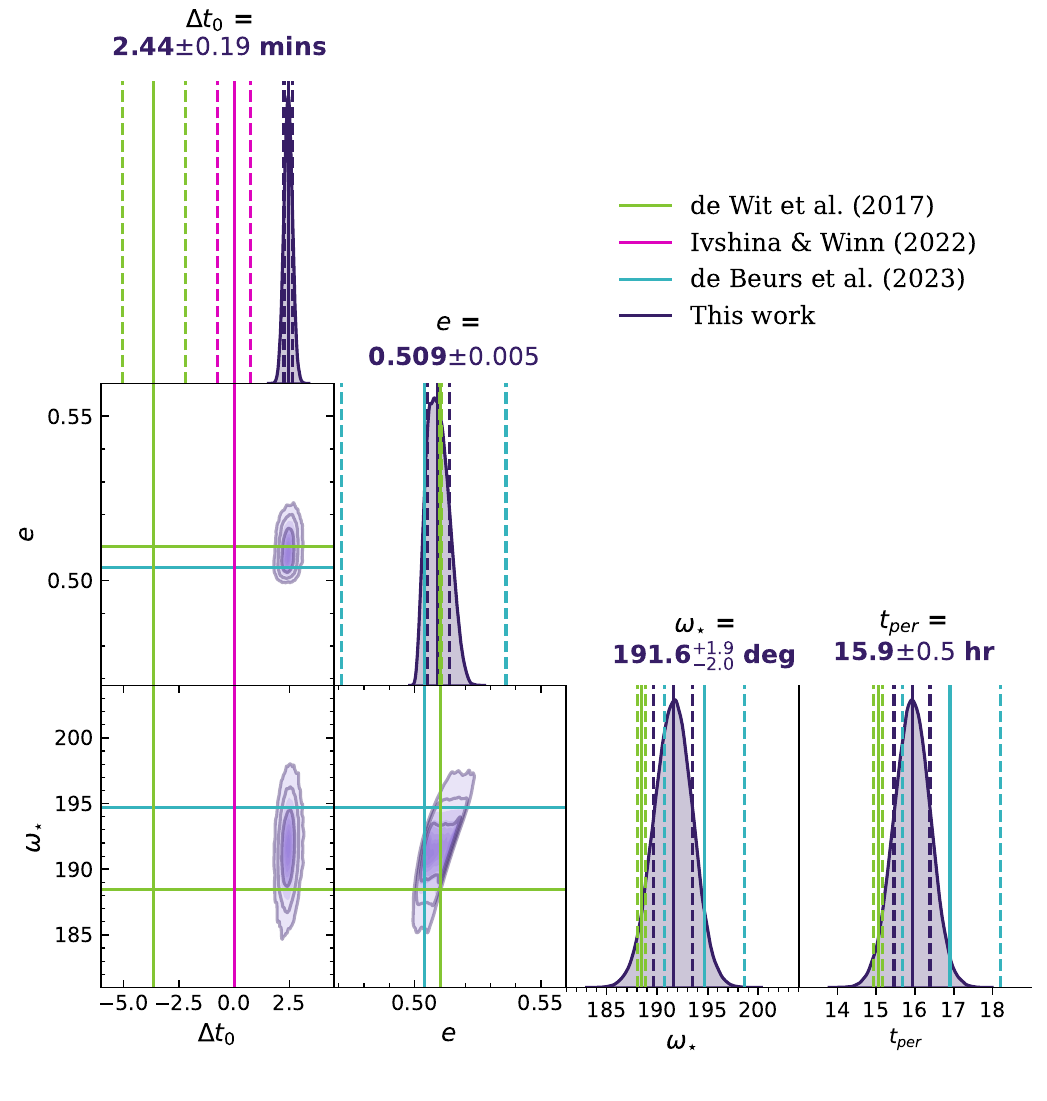}
    \caption{Corner-corner plot of HAT-P-2b's orbital parameters from the white light-curve fits. This is an excerpt from the full corner-corner plot in Figure \ref{fig:cornercorner} in Appendix \ref{app:additional_figs}. From left to right: $\Delta t_0$, the difference in minutes between the measured transit and the predicted timing of mid-transit from \citet{Ivshina2022}; $e$, the orbital eccentricity; $\omega_{\star}$, the argument of periastron; and $t_{per}$, the difference in hours between $t_0$ and the time of periastron that is derived from the measured $t_0$, $e$, and $\omega_{\star}$. Because $t_{per}$ is almost uniquely defined by $\omega_{\star}$, we do not show its correlations. Median values are denoted with a solid line in the histograms and $1\sigma$ values are denoted with dashed lines. We overplot the values found by \citet{deWit2017} (green), \citet{Ivshina2022} (brown) and \citet{deBeurs2023} (blue).
    }

    \label{fig:orbparams}
\end{figure}

In Figure \ref{fig:orbparams} we show a corner-corner plot between the most important retrieved orbital parameters and we compare them to literature values from \citet{deWit2017} and \citet{Ivshina2022}. We also include predictions from the model by \citet{deBeurs2023}. We measured an eccentricity of $e=0.509\pm0.005$ and an argument of periastron $\omega_{\star}=191.6^{+1.9}_{-2.0}$\,degrees. The first is consistent ${\leq}1\sigma$ with previous works, while $\omega_{\star}$ is 1.6\,$\sigma$ larger.

We found the transit to occur at $t_0=2459204.11219\pm0.00014$ BJD TDB, which is discrepant with \citet{Ivshina2022} at $3.3$\,$\sigma$. 
This difference is potentially due to the tidal star-planet interactions. 

From the posteriors of $\Delta t_0$, $e$, and $\omega_{\star}$, we calculated the time of secondary eclipse to be $t_{\rm{sec}}=2459205.2094\pm0.012$ BJD TDB and the time of periastron to be $t_{\rm{per}}=15.9\pm0.5$\,hr after the primary transit. The precision on $t_{\rm{sec}}$ is much lower than on $t_0$ because we did not clearly detect the ingress or egress of the secondary eclipse. The time between periastron and the primary transit was $0.67\pm0.5$\,hr longer than in the epoch \citet{deWit2017} observed (2011-2015) and consistent with the predictions from \citet{deBeurs2023} for this epoch (December 2020).

\subsection{Stellar pulsations}
We included two stellar pulsation modes. The amplitudes we measure, without having put priors on them, are $23\pm5$\,ppm and $29\pm5$\,ppm: which respectively corresponds to $0.66\pm0.14$ and $1.02^{+0.16}_{-0.17}$ times the amplitudes found by \citet{deWit2017} at $4.5$\,$\mu$m. Also, the pulsation phases are very similar to those in \citet{deWit2017}: the pulsation modes interfere destructively during transit and constructively during eclipse.

\subsection{Planetary phase curve parameters}
\label{sec:results:WL}
\begin{figure*}
    \centering
    \includegraphics[width=1.75\columnwidth]{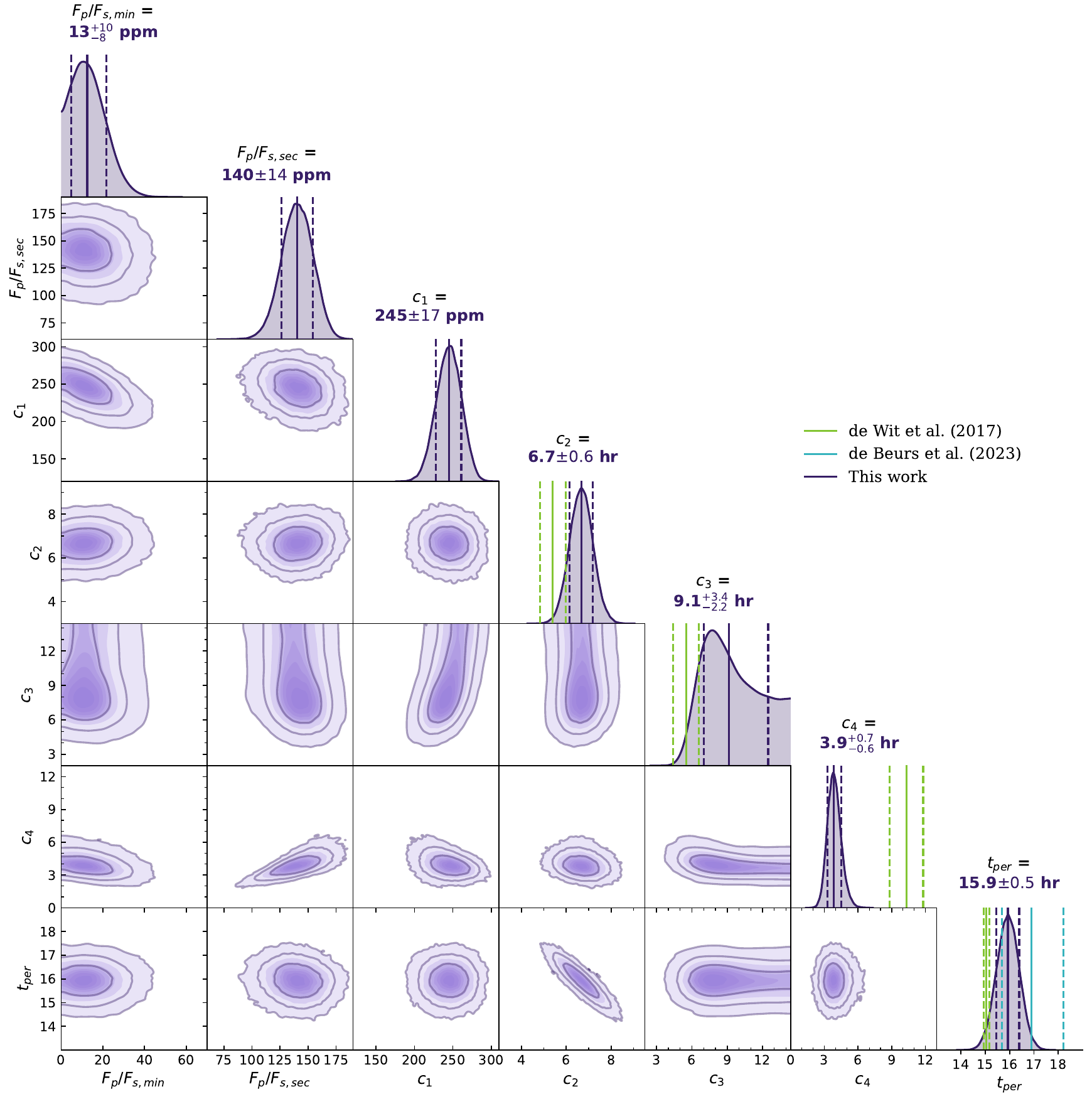}
    \caption{Corner-corner plot of HAT-P-2b's phase curve parameters from the full light-curve fits. Correlations between the fit parameters shown in this Figure and other fit parameters are shown in Figure \ref{fig:cornercorner} in Appendix \ref{app:additional_figs}. From left to right: $F_p/F_{s,min}$, the minimum planetary flux in ppm;  $F_p/F_{s,sec}$, the secondary eclipse depth in ppm, which is not fit but derived from the other parameters; $c_1$, the difference in planetary flux between its minimum and maximum in ppm; $c_2$, the difference in hours between periastron and the time of flux maximum; $c_3$, the flux increase timescale in hours; $c_4$, the flux decrease timescale in hours; and $t_{per}$, the difference in hours between $t_0$ and the time of periastron. We overplot the values found by \citet{deWit2017} (green), and \citet{deBeurs2023} (blue). Median values are denoted with a solid line in the histograms and $1$\,$\sigma$ uncertainties are denoted with dashed lines.  
    }

    \label{fig:phasecurveposteriors}
\end{figure*}

We present the posteriors of the phase curve fit parameters in Figure \ref{fig:phasecurveposteriors}. We found a peak $F_p/F_s$ of $258\pm14$\,ppm about $c_2=6.7\pm0.6$\,hr after periastron passage and $F_p/F_{s, \rm{min}} = 13^{+10}_{-8}$\,ppm. This corresponds to Earth-facing, hemisphere-averaged planetary brightness temperatures of $2360\pm30$\,K and $1390_{-160}^{+120}$\,K respectively. 
Our measurements find a timescale for planetary flux increase $c_3=9.1^{+3.4}_{-2.1}$, and a much shorter timescale for planetary flux decrease $c_4=3.9^{+0.7}_{-0.6}$\,hr.

\begin{figure*}
    \centering
    \includegraphics[width=\textwidth]{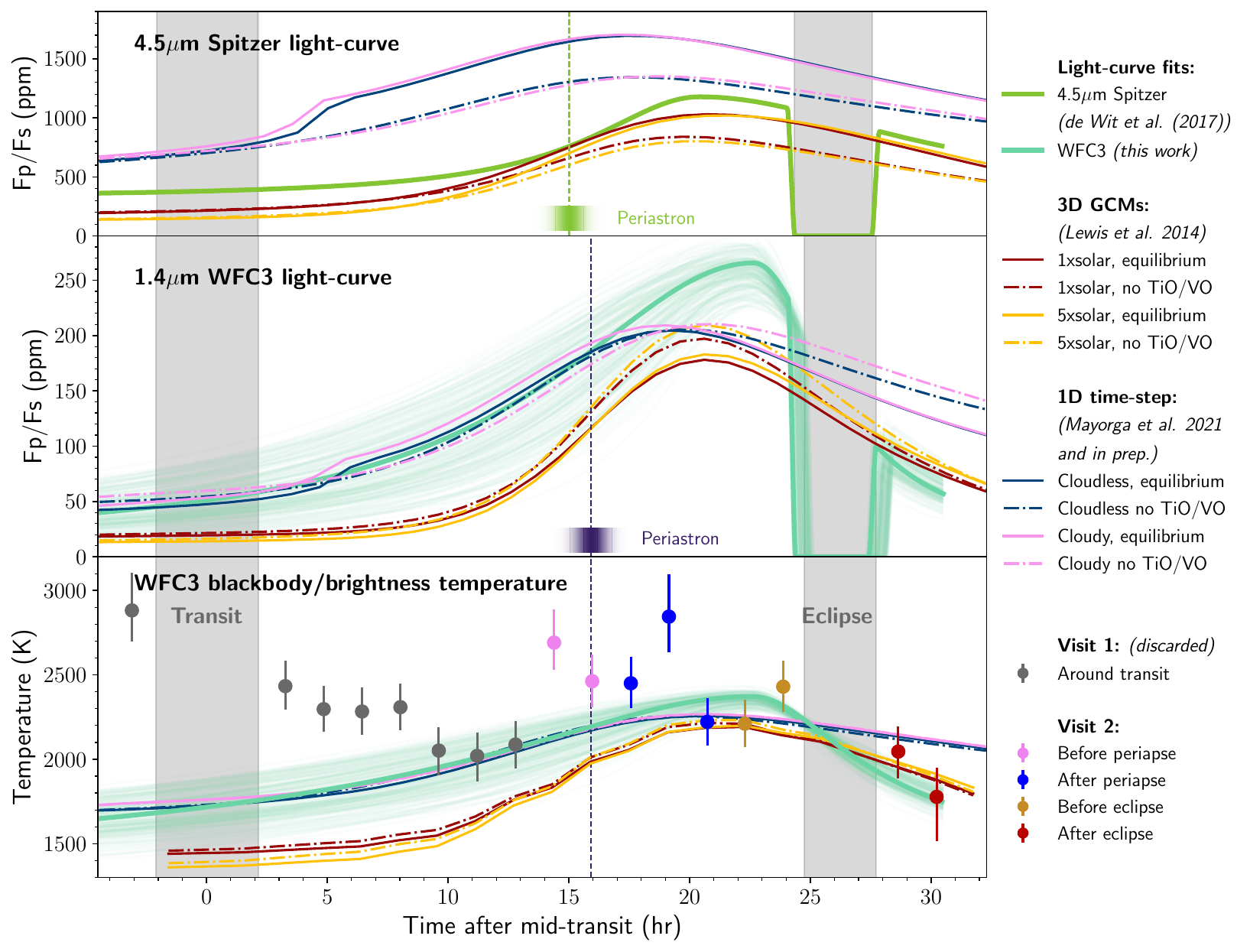}
    \caption{HAT-P-2b's partial phase curve compared to a suite of forward models. We denote the 3D GCMs by \citet{Lewis2014} in red for the solar metallicity case and in yellow for the $5\times$solar metallicity case. Cloudless 1D solar metallicity forward models by \citep[][and in prep.]{Mayorga2021} are blue and cloudy models are pink. For the models we denote the equilibrium case with a solid line and cases where TiO and VO are excluded with a dash-dotted line. We also denote the posterior distribution of the time of periastron from this work (purple) and from \citet{deWit2017} and we show their median as a dashed line. The 1D time-stepping forward models are for the planet's dayside rather than the Earth-facing hemisphere. Between periastron and secondary eclipse this is a decent approximation. \\
    \textbf{Upper:} Planetary flux relative to the stellar flux at Spitzer's $4.5$\,$\mu$m wavelengths from \citet{deWit2017} in green. The models are sampled at the epoch, and therefore orbital geometry, at which \citet{deWit2017} observed. The 1D time-step show a sudden step in brightness at ${\sim}5$ hours after mid-transit due to the sudden onset of a thermal inversion and therefore a sudden change in atmospheric layers probed. \\
    \textbf{Middle:} $F_p/F_s$ at WFC3 wavelengths. 200 posterior samples from Figure \ref{fig:rawdata} are shown in aquamarine green with the best-fit light-curve being bolded. The models are sampled at the epoch of the WFC3 observations. \\
    \textbf{Lower:} Common-mode blackbody temperature phase curve compared to modeled blackbody temperatures and brightness temperatures of the white light-curve fits. The measured temperatures were calculated for each orbit using the common-mode method and a blackbody fit. We converted the $F_p/F_s$ from the models to blackbody temperatures by fitting a blackbody to their phase-resolved spectra, hence why the GCM temperatures are more bumpy than the modeled band-averaged flux. The common-mode measured temperatures of the second visit are also noted in Table \ref{tab:cm_temperatures}. We discard the common-mode results for the first visit because of the strong telescope systematics as discussed in Section \ref{sec:analysis:phase}.}
    \label{fig:GCMmodels}
\end{figure*}

 We compare our results against the 3D GCMs as well as the 1D time-stepping models in Figure \ref{fig:GCMmodels}. Until secondary eclipse the flux is higher than the flux modeled by the GCMs, but after eclipse the flux is overestimated by the 1D model. 
 Unlike at $4.5$\,$\mu$m, the differences in expected flux between the versions of the two types of forward models is not large enough to make a significant distinction between them. 
 The differences between the cloudy and cloudless models are small because the modeled cloud layers are deeper than the WFC3 photosphere.

A crucial difference between the two categories of forward models lies in the flux decrease timescale. In the GCMs the flux decreases on a timescale of ${\sim}8$\,hours, and in the 1D time-stepping models it decreases in ${\sim}17$\,hours. The difference between the two types of forward models is largely rooted in their treatment of advection and the inherent three-dimensional nature of the planet. Contrary to the atmospheric columns modeled by the GCMs, the 1D time-stepping models have no neighboring columns for which to calculate the mutual exchange of heat. As such, the 1D time-stepping models do not consider horizontal advection. Also, the flux shown in Figure \ref{fig:GCMmodels} for the 1D time-stepping model is the calculated flux of the dayside rather than the Earth-facing hemisphere. Between periastron and secondary eclipse this is a decent approximation since the flux emanating from the Earth-facing hemisphere is dayside-dominated at those times. However, accounting for rotation and therefore the visible fraction of the dayside, would decrease the flux increase/decrease timescales and bring them more in line with the 3D GCMs and the observed flux.

\subsection{Transit and eclipse spectra}
\label{sec:results:spectra}
\begin{figure}
    \centering
    \includegraphics[width=\columnwidth]{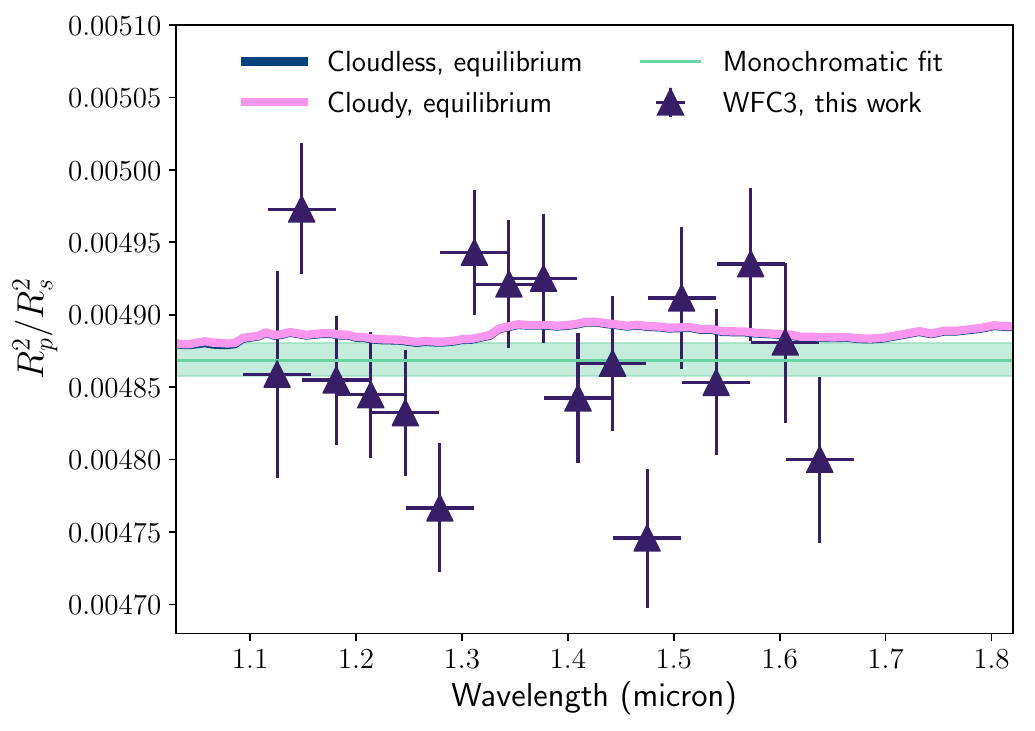}
    \caption{HAT-P-2b's transmission spectrum measured for WFC3. We overplot the monochromatic fit (aquamarine green) as well as cloudless (blue) and cloudy (pink) 1D equilibrium chemistry models from Section \ref{sec:1D_time_step}. The shaded region on the fit denotes a $1\sigma$ uncertainty on the wavelength-averaged transit depth. Because the clouds in the cloudy model reside at very high pressures, the cloudless and cloudy models are so similar that they are almost indistinguishable in this plot.}
    \label{fig:transmission-spectrum}
\end{figure}

In Figure \ref{fig:transmission-spectrum} we exhibit the WFC3 transmission spectrum. The atmospheric scale height during transit is approximately 24\,km given a temperature of 1300\,K and a mean molecular mass of 2.33 \citep{Lewis2014}. This corresponds to 3\,ppm in terms of $R_p^2/R_s^2$, the transit depth. Because the mean spectral uncertainty for the transit spectrum is 49\,ppm we can draw no conclusions on the atmospheric condition during transit. This notion is corroborated by the forward 1D equilibrium chemistry models.
Indeed, the WFC3 spectrum has no detected slope within $0.5$\,$\sigma$. However, its reduced $\chi^2$ is $1.8$, indicating that the scatter is larger than expected, largely due to the telescope systematics ($\delta_\lambda$).

As the planet transits in front of the star, its nightside is visible to the observer, carrying an imprint on the transmission spectrum. Due to the nature of the fitting method used for the transit data, we are insensitive to the nightside flux itself and we can only measure its variation. From the fitted light-curve we estimate that, during transit, the planet's Earth-facing brightness temperature increased by ${\sim}30$\,K, introducing a potential slope of ${\sim}10$\,ppm\,$\mu$m$^{-1}$. This is much smaller than the measured uncertainty on this slope of $40$\,ppm\,$\mu$m$^{-1}$. Changes in spectral features would be even smaller. Planetary nightside flux variations during transit can therefore safely be ignored for the transmission spectrum.

Because the expected spectral transmission features are so small, we assume the spectrum to be featureless. WFC3's precision therefore allows us to improve the precision on the relative planetary radius by a factor 3 with respect to the previous measurement by \citet{deWit2017}.  
At transit we measure a WFC3 weighted average planetary radius over stellar radius of $R_p^2/R_s^2 = 0.004869 \pm 0.000012$. 

\begin{figure*}
    \centering
    \includegraphics[width=\textwidth]{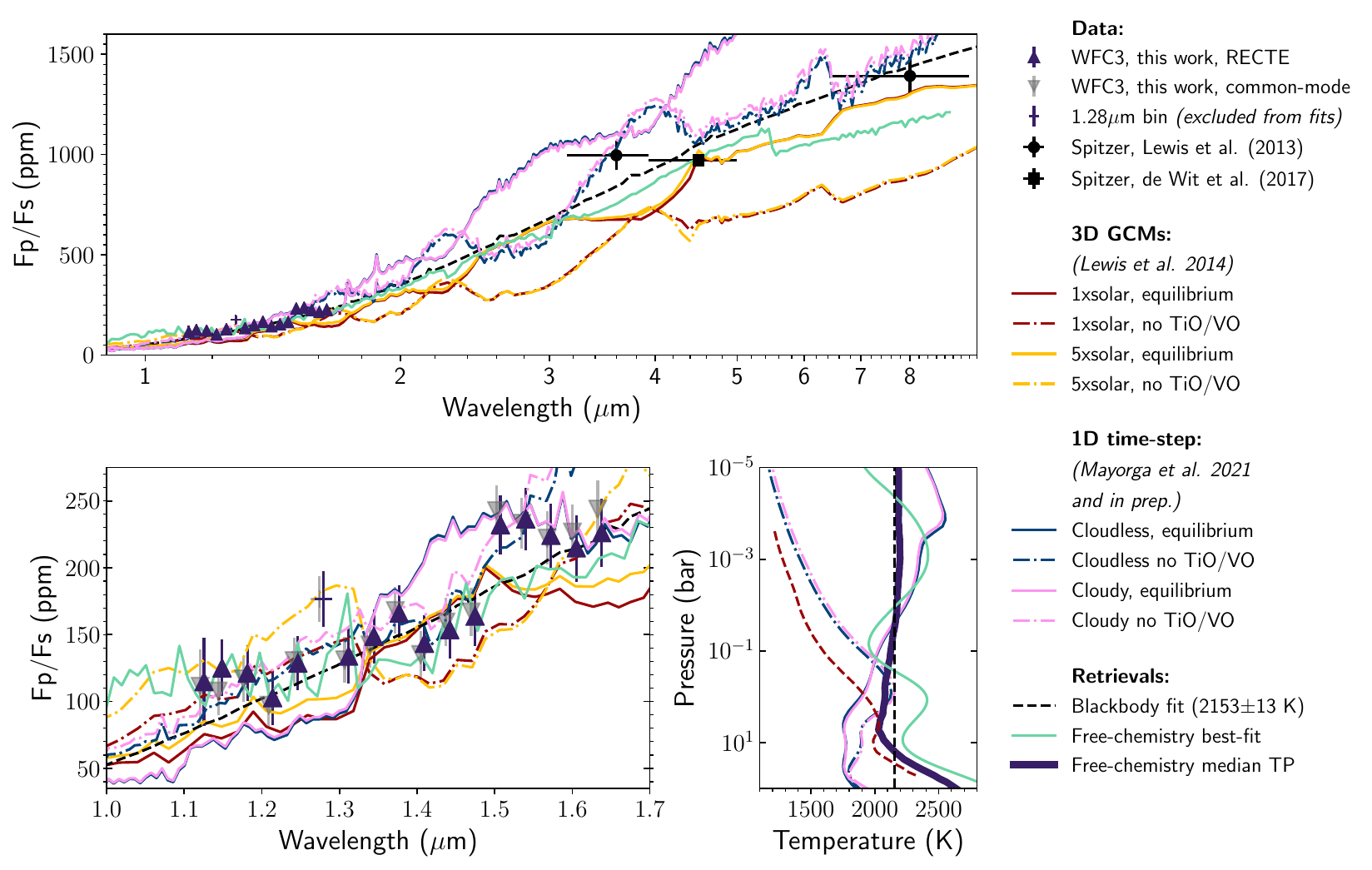}
    \caption{HAT-P-2b's eclipse spectrum for WFC3 with the RECTE fit (purple), with the common-mode method (grey) and for the Spitzer observations (black) \citep{Lewis2013, deWit2017}. We also overplot a blackbody fit (dashed black line), the 3D GCM models (red, yellow), the 1D time-stepping models (blue, pink), and the best-fit 1D free-chemistry retrieval (aquamarine green). The retrieval was performed on the WFC3 data only, but excluding the bin centered at $1.28$\,$\mu$m (denoted with a cross). See Section \ref{sec:results:Passchen} for a discussion on this.\\
    \textbf{Upper:} Full HAT-P-2b near-infrared eclipse spectrum. We sampled the forward models at what was the time of secondary eclipse during the epoch of the Spitzer observations (2011). This makes for a ${\sim}80$\,ppm difference at Spitzer wavelengths compared to the WFC3 epoch (2020). The Spitzer data by \citet{Lewis2013} do not take any stellar pulsations into account.\\
    \textbf{Lower left:} Eclipse spectrum zoomed in on the WFC3 wavelengths. The models are sampled at the epoch of the WFC3 observations (2020). We added a monochromatic offset to the common-mode reduced spectrum to match the average eclipse depth of the RECTE fit. We also offset the common-mode reduced spectrum in wavelength slightly to make it more distinguishable from the RECTE fit. Note that the common-mode method is entirely different from the \texttt{RECTE} method, yet the results are relatively similar.\\
    \textbf{Lower right:} Temperature pressure profiles of the models, sampled at the time of secondary eclipse during the WFC3 epoch. The equilibrium forward models have a thermal inversion, while the models without TiO/VO do not. The median TP-profile (purple) of the free-chemistry retrieval is almost isothermal, while the best-fit profile (aquamarine green) is not.}
    \label{fig:eclipse-spectrum}
\end{figure*}

Figure \ref{fig:eclipse-spectrum} displays the secondary eclipse spectrum from the WFC3 data augmented with the Spitzer data from \citet{Lewis2013} (3.6 and 8.0\,$\mu$m) and from \citet{deWit2017} (4.5\,$\mu$m). These data, and likewise the spectral transit depths, are listed as well in Table \ref{app:tab:transit} in Appendix \ref{app:additional_figs}. One should use caution when comparing the eclipse depths of this work to literature values, because they were taken at different epochs. Therefore, they were taken with different observing geometries, and consequently with different eclipse timings compared to periapse, and therefore a different expected temperature. We excluded the Spitzer data at 5.8\,$\mu$m from \citet{Lewis2013} because eclipse measurements at those wavelengths are often found to be less reliable because of the differing telescope characteristics \citep{Desert2009, Stevenson2010, Desert2011b, Lewis2013}.

We also added forward spectral models from the 3D GCMs and the 1D time-stepping models to Figure \ref{fig:eclipse-spectrum}. Due to the different expected temperatures (see Figure \ref{fig:GCMmodels}), it is difficult to directly compare the models to the data. However, the shape of the WFC3 eclipse spectrum most closely resembles either a blackbody spectrum, or a spectrum without TiO/VO.

The lower panel of Figure \ref{fig:eclipse-spectrum} compares the HAT-P-2b WFC3 spectrum eclipse spectrum from the \texttt{RECTE} method (purple) to the common-mode reduced spectrum (grey). For the common-mode based spectrum we used the same orbits as for the \texttt{RECTE} fit and we added a monochromatic constant such that the means of both spectra are the same. Despite a visible difference in slope, the two spectra are in decent agreement with each other and differ by at most 0.8\,$\sigma$.
We stress here that these methods are based on completely different principles, yet they produce a similar spectrum. This is a testament to the possibilities that the common-mode based method by \citet{Arcangeli2021} creates.

The $F_p/F_s$ of the \texttt{RECTE} method in the spectral bin centered at $1.28$\,$\mu$m is $2.4\sigma$ higher than the blackbody fit and it does not coincide with any spectral features in the models. By itself, such deviation does not warrant this bin to be flagged as anomalous, but the deviation is also visible throughout the partial phase curve (see Section \ref{sec:results:Passchen}). We therefore exclude this bin from the spectral retrievals and discuss its origin in Section \ref{sec:discussion:Paschen}. Excluding this spectral bin, a blackbody fits the WFC3 data at $\chi_\nu^2=0.83$.

In Appendix \ref{app:Changeat_Edwards} we compare the transmission and eclipse spectra to those found by \citet{Changeat2022} and \cite{Edwards2022}.

\subsection{Spectral retrievals}
\label{sec:results:retrieval}
We used the 1D free-chemistry retrievals outlined in Section \ref{sec:1D_free-chemistry_models} to chemically and thermally characterize HAT-P-2b's atmosphere during eclipse. The resulting best-fit is depicted in aquamarine green in Figure \ref{fig:eclipse-spectrum} with the corresponding corner-corner plot shown in Figure \ref{fig:eclipse_retrieval_corner} in the Appendix. In Figure \ref{fig:contrib_fct} we plot the best-fit TP-profile with the corresponding contribution function. It is notable that the median TP-profile looks nearly isothermal, while the best-fit model is not. This indicates multiple possible solutions to the spectrum. The contribution function of the best-fit model indicates that we probe lower pressures than the forward models predict.
Contrary to the expectations based on the spectra of the forward models, we retrieve a high median H$^-$ abundance of log(H$^-$)$=-4.0^{+2.3}_{-5.4}$ while the TiO and H$_2$O abundances are largely unconstrained within our priors. 

\begin{figure}
    \centering
    \includegraphics[width=\columnwidth]{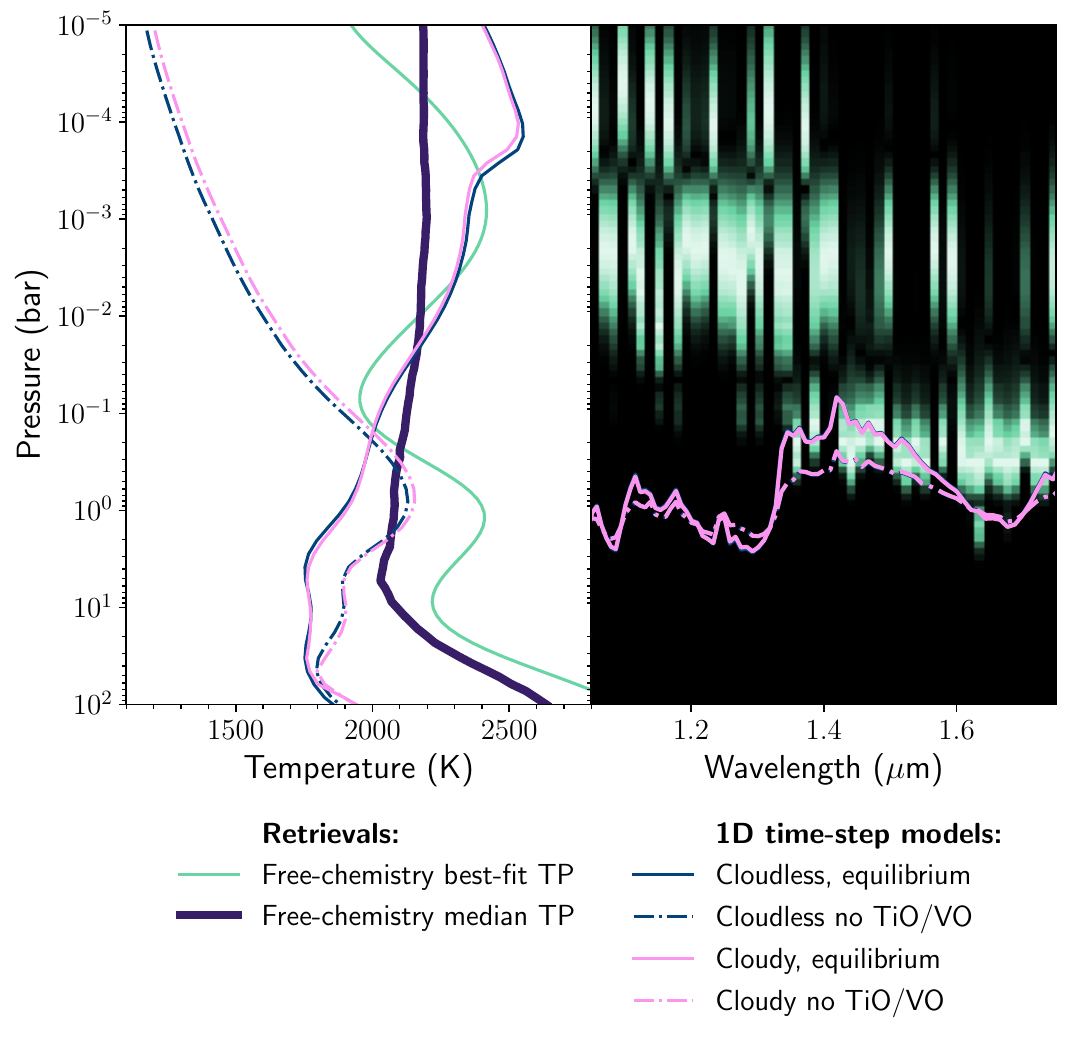}
    \caption{Best-fit (aquamarine green) and median (purple) temperature pressure (TP) profiles of the 1D free-chemistry retrievals of the secondary eclipse spectrum. The panel on the right displays the contribution function that corresponds to the best-fit TP-profile. We compare this to the TP-profiles and the pressure of the $\tau=0.5$ layer of the 1D time-stepping forward models. There are almost no differences between the cloudy and cloudless contribution functions.}
    \label{fig:contrib_fct}
\end{figure}

The 1.28\,$\mu$m bin not only coincides with the Pa$\beta$ line, but it is also close to the peak of a TiO opacity band. Therefore, if the 1.28\,$\mu$m bin had been included in the fits, the TiO volume mixing ratio would run up against the edge of the prior, although low TiO abundances are not excluded. 

In order to test and quantify the significance of the presence of the molecules detected in the secondary eclipse spectrum, we ran identical retrievals with each molecular opacity source removed. By comparing the Bayesian evidences, we calculate an equivalent detection significance of each molecule. We list the results in Table \ref{tab:bayesian_evidence} and we conclude that the presence of TiO and H$_2$O are the most significant at 1.5\,$\sigma$. However, neither makes the 3$\sigma$ threshold to qualify for a detection.

\section{Phase-resolved spectra results}
\label{sec:results:partialphase}
\subsection{Blackbody temperature fits}
\label{sec:results:partialphase:blackbody}
To investigate the evolution of HAT-P-2b's spectrum throughout its orbit, we utilized the common-mode method discussed in Section \ref{sec:analysis:phase}. We fitted a blackbody model to the spectra using an MCMC method \citep{emcee} to retrieve a phase-resolved blackbody temperature. Importantly, the common-mode method produces planetary emission spectra that are only \emph{relative} up to a monochromatic constant. We therefore added a monochromatic constant to fits of the common-mode spectra. We used a PHOENIX stellar model for HAT-P-2 \citep{Husser2013}, and we again excluded the wavelength bin centered around 1.28\,$\mu$m.

We present the retrieved blackbody temperatures in Figure \ref{fig:GCMmodels} and compare them to a brightness temperature curve derived from the white light-curve fits by multiplying them with a PHOENIX stellar models \citep{Husser2013}.
In order to also properly compare the performance of the common-mode method to the predictions, we fit a blackbody to the 3D and 1D model spectra at WFC3 wavelengths for each time stamp. The forward models show a significantly lower planetary temperature, especially for the planet's nightside during transit. Close to secondary eclipse, the retrieved temperatures align very well with the 3D models and the light-curve-derived temperature curve, but earlier orbits stray away from them.  In the blackbody fits to the forward models we did not fit for any monochromatic offset. Including such offset would significantly decrease the modeled blackbody temperature, further increasing the difference with the data.

Noticeably, there is a clear step between the first and second visit. We emphasize that the common-mode based method investigates the spectrum of an HST orbit with respect to the spectrum in secondary eclipse. Therefore, the more similar the telescope systematics and stellar spectrum are to those in eclipse, the more accurate the observed planetary phase-resolved spectra will be. In Section \ref{sec:analysis:phase} we established that the telescope systematics are too large for the first visit, but possibly acceptable for the second visit. Henceforth we will concentrate on the phase-resolved spectra of the second visit and ignore the first visit's phase-resolved spectra. We tried recovering useful spectra from the first visit by dividing those spectra by the two orbits just after transit instead of the orbits in eclipse. However, those first-visit common-mode-reduced spectra still showed significant systematics and their slopes were either too steep (indicating an extremely hot planet), or negative.

We suggested in Section \ref{sec:analysis:phase} that existing telescope systematics and/or stellar effects may result in the planetary temperature during the first orbit of the second visit to be underestimated by ${\sim}1800$\,K. We find the contrary: the measured temperatures at the start of the second visit are higher than expected.

Table \ref{tab:cm_temperatures} lists the retrieved blackbody temperatures and offsets, and Figure \ref{fig:phase} shows the phase-resolved spectra and the blackbody spectra. We combined the orbits into four groups: the two orbits before periastron (pink), the three orbits just after periastron (blue), the two orbits just before eclipse (gold) and the two orbits after eclipse (red). We also show blackbody fits with priors on the monochromatic offset. We uitilized the posterior distribution of $F_p/F_s$ from the white light-curve fit as a prior on the monochromatic offset. These blackbody fits show that slightly more sensible results can be obtained if the light-curve is known. However, they are still very different from the forward models around periapse.

Figures \ref{fig:phase}, \ref{fig:TPs} and Table \ref{tab:cm_temperatures} suggests that the spectral temperature declines throughout the visit, albeit not significantly until after secondary eclipse. In Section \ref{sec:results:partialphase:stellar} we assess what the stellar contribution to this decline could be.

The spectral structure observed in the eclipse spectrum in Section \ref{sec:results:spectra} is also visible in the phase-resolved spectra in Figure \ref{fig:phase}, indicating that the chemistry of the planet's atmosphere at the pressures probed by this observation did not change significantly throughout this phase of the planetary orbit. Moreover, the spectral bin at $1.28$\,$\mu$m is consistently more than $1\sigma$ above the blackbody fit. We further discuss this feature in Section \ref{sec:results:Passchen}. 

\subsection{Spectral retrievals}
\label{sec:results:partialphase:retrievals}
\begin{figure*}
    \centering
    \includegraphics[width=\textwidth]{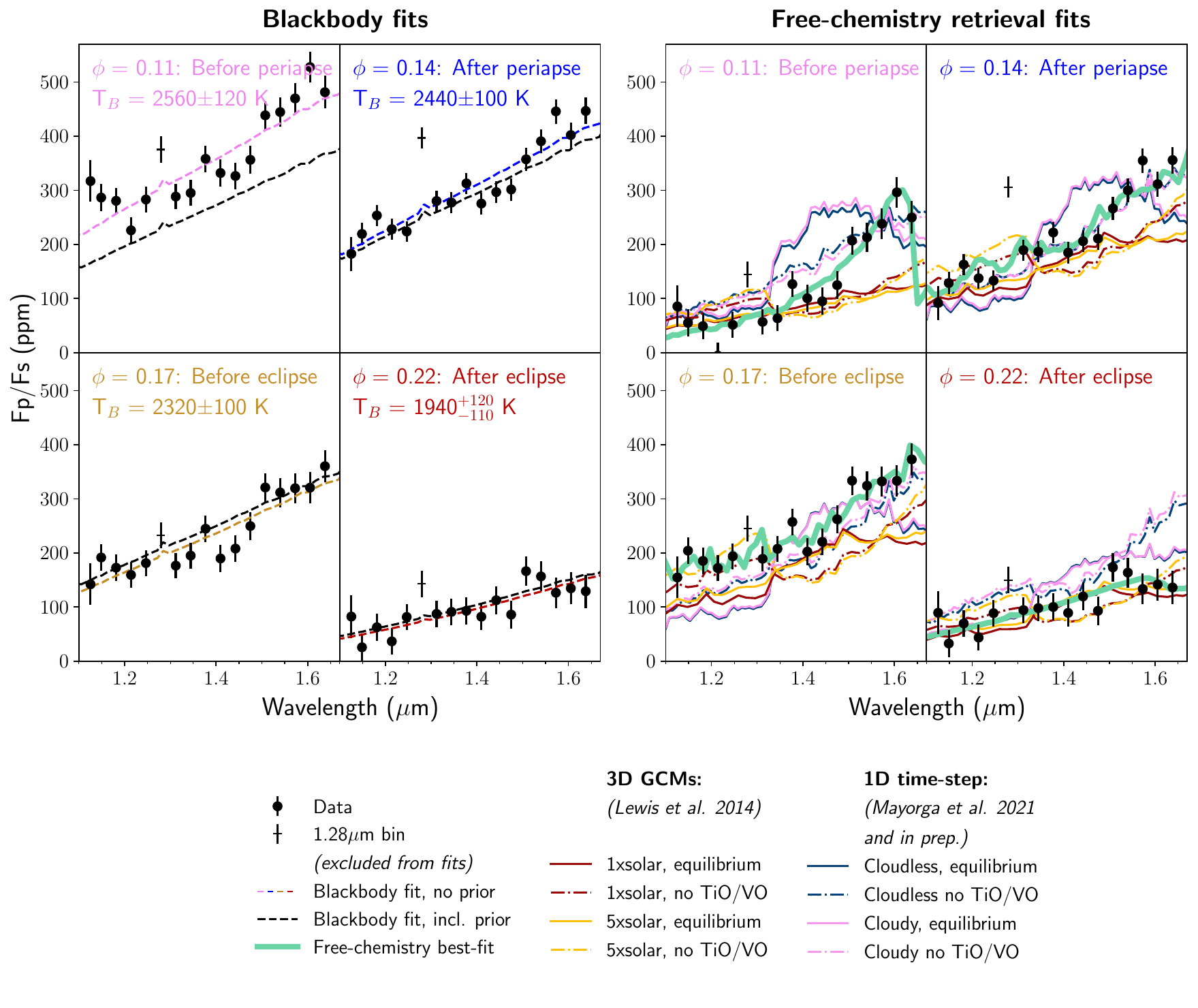}
    \caption{Partial phase curve spectra for HAT-P-2b of the second visit for the common-mode method (black data) compared to forward models and the 1D free-chemistry retrieval (aquamarine green). The 3D GCMs are represented in red for the solar metallicity scenario and in yellow for the $5\times$solar metallicity scenario. 1D cloudless solar metallicity forward models by \citep[][and in prep.]{Mayorga2021} are depicted in blue, while cloudy models are shown in pink. Equilibrium cases are denoted with solid lines, and cases excluding TiO and VO are represented with dash-dotted lines.
    Dashed lines represent fits of a blackbody spectrum plus a monochromatic offset. The fits without priors are colored and the fits including temperature priors from the white light-curve, are black. 
    Any features in the blackbody fits are of stellar origin. We excluded the wavelength bin at 1.28\,$\mu$m from the fits. 
    In the four panels on the left, the common-mode spectra have been shifted with a monochromatic offset that fits the blackbody spectra without priors. In the four panels on the right they are shifted to best fit the 1D free-chemistry retrievals. Due to the nature of the common-mode method, these offsets are not known \textit{a priori}. The big difference in monochromatic offset between the blackbody fits and the free-chemistry retrievals is readily visible and quantified in Table \ref{tab:cm_temperatures}. 
    }
    \label{fig:phase}
\end{figure*}

In Section \ref{sec:results:retrieval} we performed 1D free-chemistry retrievals on the secondary eclipse spectrum. We performed similar retrievals on the partial-phase curve spectra, again excluding the bin at 1.28\,$\mu$m. However, due to the nature of the common-mode method, we added an extra parameter to fit for the monochromatic offset. We again leveraged the posterior distribution of $F_p/F_s$ from the white light-curve fit as a prior to the offset. 
In Figure \ref{fig:phase} we plot the retrieved spectra as aquamarine green lines. In Table \ref{tab:bayesian_evidence} we provide the retrieved volume mixing ratios and the statistical significance of the detection of each molecule. In Figure \ref{fig:contrib_fct_phase} in the Appendix we plot the contribution functions corresponding to the best fits.

\begin{table*}[]
\centering
\begin{tabular}{@{}l@{}||lll|lll|lll|lll|lll}
Phase & \multicolumn{3}{c|}{H$_2$O} & \multicolumn{3}{c|}{TiO} & \multicolumn{3}{c|}{VO} & \multicolumn{3}{c|}{H$^{-}$} & \multicolumn{3}{c}{FeH}\\
\cline{2-16}
& log(X$_i$) & $\mathcal{B}$ & Sign. & log(X$_i$) & $\mathcal{B}$ & Sign. & log(X$_i$) & $\mathcal{B}$ & Sign. & log(X$_i$) & $\mathcal{B}$ & Sign. & log(X$_i$) & $\mathcal{B}$ & Sign.\\
\hline
\hline
\cellcolor{myviolet!40}Before periapse& $-6.8^{+3.3}_{-3.2}$ & 0.8 & 1.0\,$\sigma$ & $-6.5^{+3.3}_{-3.4}$ & 1.1 & 1.1\,$\sigma$ & $-6.5^{+3.1}_{-3.4}$ & 0.9 & 0.9\,$\sigma$ & $-4.2^{+0.4}_{-1.6}$ & 4.5 & 2.3\,$\sigma$ & $-5.6^{+2.8}_{-3.8}$ & 1.3 & 1.5\,$\sigma$ \\
\cellcolor{myblue!40}After periapse& $-6.3^{+3.5}_{-3.5}$ & 1.0 & 1.1\,$\sigma$ & $-5.3^{+2.7}_{-4.0}$ & 1.2 & 1.4\,$\sigma$ & $-3.0^{+0.8}_{-4.9}$ & 0.8 & 1.0\,$\sigma$ & $-6.5^{+2.5}_{-3.2}$ & 0.6 & 1.1\,$\sigma$ & $-6.6^{+3.0}_{-3.2}$ & 1.0 & 0.9\,$\sigma$ \\
\cellcolor{mylime!40}Before eclipse& $-6.0^{+3.6}_{-3.5}$ & 1.3 & 1.4\,$\sigma$ & $-5.7^{+3.7}_{-3.7}$ & 1.0 & 0.9\,$\sigma$ & $-6.1^{+3.3}_{-3.5}$ & 0.9 & 0.9\,$\sigma$ & $-4.5^{+2.6}_{-5.0}$ & 0.5 & 1.2\,$\sigma$ & $-5.2^{+2.6}_{-3.8}$ & 1.5 & 1.6\,$\sigma$ \\
\cellcolor{mypurple!40}Secondary eclipse& $-5.8^{+3.5}_{-3.9}$ & 1.3 & 1.5\,$\sigma$ & $-6.2^{+3.1}_{-3.3}$ & 1.4 & 1.5\,$\sigma$ & $-6.8^{+3.5}_{-3.1}$ & 1.2 & 1.3\,$\sigma$ & $-4.0^{+2.3}_{-5.4}$ & 0.9 & 0.9\,$\sigma$ & $-5.0^{+2.7}_{-4.0}$ & 1.2 & 1.3\,$\sigma$ \\
\cellcolor{myred!40}After eclipse& $-6.1^{+3.4}_{-3.5}$ & 1.1 & 1.3\,$\sigma$ & $-6.5^{+3.3}_{-3.3}$ & 1.1 & 1.2\,$\sigma$ & $-6.6^{+3.4}_{-3.4}$ & 1.2 & 1.3\,$\sigma$ & $-4.6^{+2.7}_{-4.8}$ & 1.1 & 1.2\,$\sigma$ & $-5.5^{+3.0}_{-3.7}$ & 1.1 & 1.3\,$\sigma$ \\
\end{tabular}

\caption{Chemistry results of the 1D free-chemistry retrievals. For each molecule and for each partial-phase curve spectrum we list the volume mixing ratio (log($X_i$)), the Bayes factor $\mathcal{B}$ between a model with and without that molecule, as well as the equivalent statistical significance of the detection of that molecule.}
\label{tab:bayesian_evidence}
\end{table*}

\begin{figure}
    \centering
    \includegraphics[width=\columnwidth]{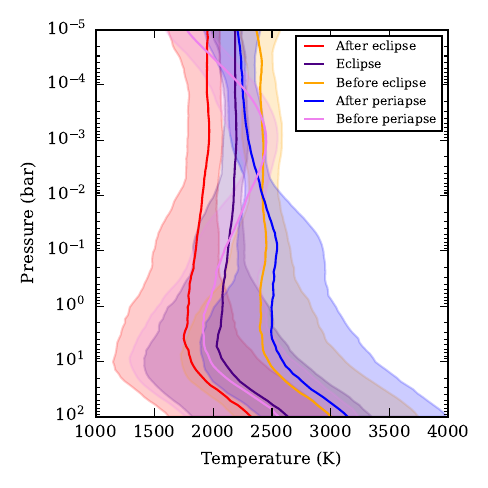}
    \caption{Median retrieved temperature-pressure profiles for the partial phase curve spectra. The shaded regions denote the 1$\sigma$ uncertainties. The solid lines are median TP-profiles.  The spectra are shown in Figure \ref{fig:phase}.}
    \label{fig:TPs}
\end{figure}

The median retrieved temperature-pressure (TP) profiles shown in Figure \ref{fig:TPs} show an evident temporal structure: the first observed phase (before periapse) has a significant thermal inversion. The disappearance of this inversion, and the hotter WFC3 photosphere after periapse would suggest that heat is deposited further down in the atmosphere. As the planet recedes further from its host star, it cools down, until after eclipse the photospheric temperature drops below 2000\,K. However, it is important to note that the retrieved WFC3 photospheric temperature is strongly driven by the priors on the monochromatic offset. As such, the significance of Figure \ref{fig:TPs} is in the vertical thermal structure rather than the absolute temperature.

The need for an extra opacity in the before-periapse spectrum is evident from the large difference between the blackbody fit without priors and the forward models in Figure \ref{fig:phase}. The fit without priors effectively probes the slope of the spectrum, which is steeper than expected from the brightness temperature of the forward models. Therefore, an extra opacity source that has a strongly positive slope throughout the WFC3 spectrum is necessary to bring down the retrieved temperature. H$^{-}$ in absorption with a temperature that decreases steeply with altitude, between $10^{-3}-10^{-5}$\,bar fits this bill. However, this also requires the layers below the WFC3 photosphere to be inverted. This provides a window into deeper, cooler layers, that could fit the relatively cool redmost data point. However, none of that is predicted by the forward models, nor has H$^{-}$ in absorption been predicted by any GCMs \citep{Parmentier2018}. Alternatively, TiO in absorption with an unrealistically high log(VMR) $\approx -2$ may also explain this spectrum, but struggles to fit for the redmost data point. 

As such, the retrieved abundances strongly depend on the slope of the spectra. However, the accuracy of these slopes remains uncertain (see Figure \ref{fig:GCMmodels} and Sections \ref{sec:analysis:phase}, \ref{sec:results:partialphase:blackbody}, and \ref{sec:results:partialphase:stellar}).

\begin{table*}[]
\centering
\begin{tabular}{l||c|cc|cc|cc|c|c}
& \multicolumn{7}{c|}{Blackbody fits} & \multicolumn{2}{c}{Free-chemistry retrieval fit} \\
Phase & BB Temperature & \multicolumn{2}{c|}{Grouped BB temperature} &\multicolumn{2}{c|}{Constant offset (ppm)} & \multicolumn{2}{c|}{$\chi^2_\nu$} & Constant  & $\chi^2_\nu$\\
& & no prior & with prior & no prior & with prior & no prior & with prior  & offset (ppm) & \\
\hline
\hline
\cellcolor{myviolet!40} 0.106 & $2690^{+200}_{-170}$ K   & \multirow{2}{*}{$2560\pm120$ K} & \multirow{2}{*}{$2320^{+50}_{-60}$ K} & \multirow{2}{*}{$350\pm70$} & \multirow{2}{*}{$232\pm20$}& \multirow{2}{*}{$2.5$} & \multirow{2}{*}{$2.9$} &\multirow{2}{*}{$166^{+22}_{-19}$} &\multirow{2}{*}{$4.9$}\\
\cline{1-2}
\cellcolor{myviolet!40} 0.118 & $2460\pm160$ K   &                                                                    &  & &&&  &&                                                                            \\
\hline

\cellcolor{myblue!40} 0.130 & $2450^{+160}_{-150}$ K   & \multirow{3}{*}{$2470\pm100$ K}  & \multirow{3}{*}{$2400^{+50}_{-60}$ K}                              &\multirow{3}{*}{$291\pm50$} &\multirow{3}{*}{$274^{+19}_{-24}$}& \multirow{3}{*}{$1.9$} & \multirow{3}{*}{$2.0$} &\multirow{3}{*}{$224^{+25}_{-26}$} &\multirow{3}{*}{$4.7$}\\
\cline{1-2}
\cellcolor{myblue!40} 0.142 & $2840^{+260}_{-220}$ K   &                                                                    &                                                                                 & &&&&&\\
\cline{1-2}

\cellcolor{myblue!40} 0.153 & $2220^{+150}_{-140}$ K   &                                                                    &                                                                                 & &&&&&\\
\hline
\cellcolor{mylime!40} 0.165 & $2210\pm150$ K   & \multirow{2}{*}{$2320\pm100$ K} & \multirow{2}{*}{$2390^{+40}_{-50}$ K}                                               &\multirow{2}{*}{$230\pm50$} &\multirow{2}{*}{$255\pm26$}& \multirow{2}{*}{$1.4$} & \multirow{2}{*}{$1.5$} &\multirow{2}{*}{$250^{+15}_{-16}$} &\multirow{2}{*}{$3.8$}\\
\cline{1-2}
\cellcolor{mylime!40} 0.177 & $2430\pm160$ K   &                                                                    &                                                                                 & &&&&&\\
\hline
\cellcolor{myred!40} 0.212 & $2050^{+150}_{-160}$ K    & \multirow{2}{*}{$1940^{+120}_{-110}$ K}  & \multirow{2}{*}{$1980^{+70}_{-80}$ K}                          & \multirow{2}{*}{$100^{+30}_{-40}$} &\multirow{2}{*}{$118\pm17$}& \multirow{2}{*}{$0.7$} & \multirow{2}{*}{$0.9$} &\multirow{2}{*}{$114\pm16$} & \multirow{2}{*}{$2.9$}\\
\cline{1-2}
\cellcolor{myred!40} 0.224 & $1780^{+180}_{-270}$ K   &                                                                    & & &&&&&\\
\hline

\end{tabular}

\caption{The common-mode derived blackbody temperatures of HAT-P-2b in the HST WFC3 G141 bandpass for the planetary orbital phases during the second HST visit. These temperatures are also plotted in Figure \ref{fig:GCMmodels}. In the third through tenth columns we group the phases together like in Figure \ref{fig:phase}. The third column denotes retrieved temperatures without priors. For the fourth column we included a prior on the spectral monochromatic offset from the white light-curve. The fifth, sixth and ninth columns show the monochromatic offset to be applied to the common-mode retrieved spectra. Because they describe the variations in the detected white-light, which includes telescope systematics, these offsets should be compared between methods rather than between phases. In the seventh, eighth and tenth column we denote the reduced chi-squared values of the fits. The blackbody fits perform markedly better, but as noted in the text, they are still unreliable.}
\label{tab:cm_temperatures}
\end{table*}

\subsection{Significant deviation from a blackbody at 1.28\,$\mu$m}
\label{sec:results:Passchen}
Figure \ref{fig:phase} shows that the wavelength bin centered around 1.28\,$\mu$m has an anomalously high flux compared to blackbody fits throughout the entire second visit. This bin wholly contains the strong stellar Paschen $\beta$ line at 1.282\,$\mu$m, which arises from the electronic transition within hydrogen atoms between the third energy level and the fifth. It is the most prominent stellar line at G141 wavelengths.

\begin{figure*}
    \centering
    \includegraphics[width=\textwidth]{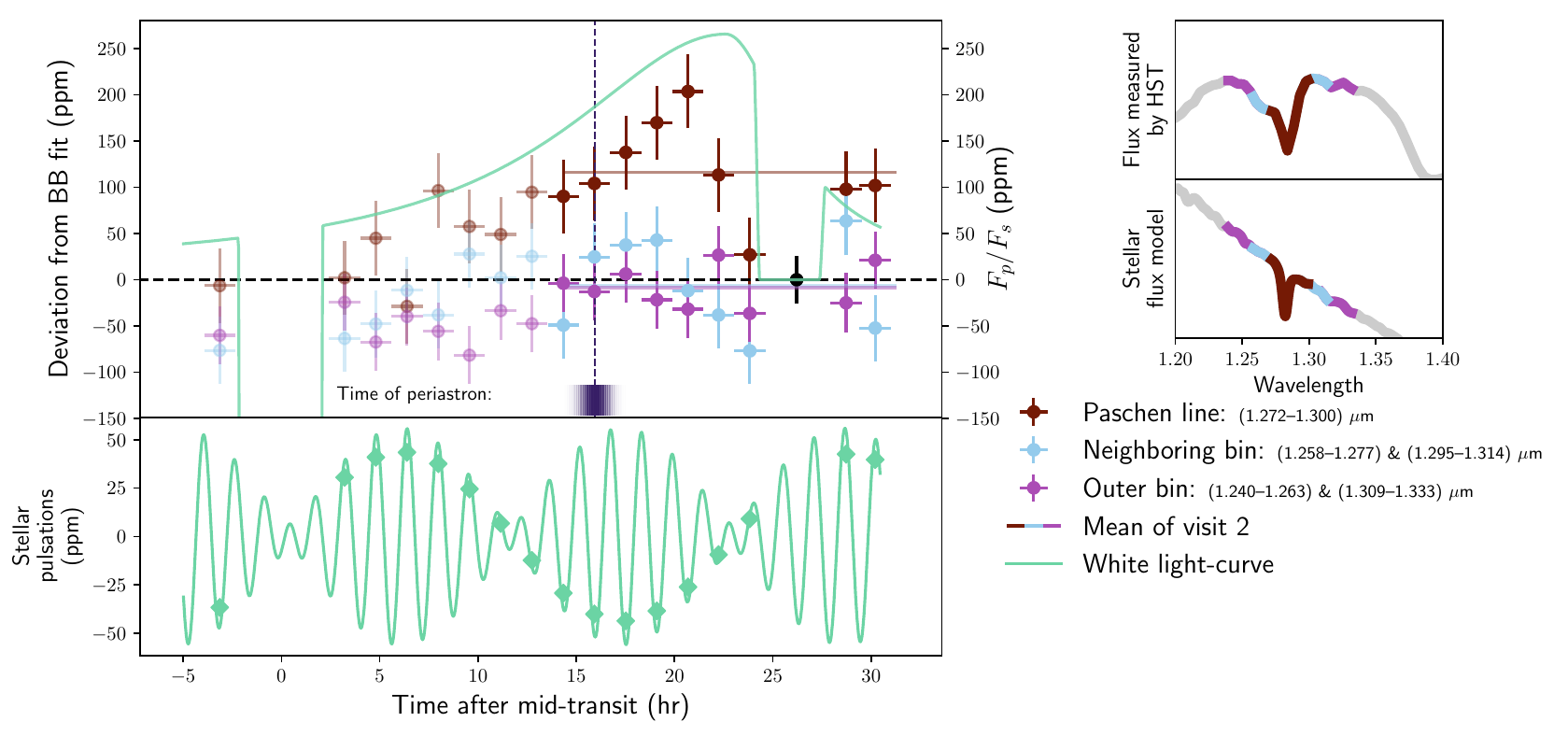}
    \caption{Excess flux of the common-mode reduced spectra at 1.28\,$\mu$m.\\
    \textbf{Left:} Deviation from a planetary blackbody model for three wavelength bins surrounding the stellar Pa$\beta$ line. For the dark red bin this equals $\delta_{\rm Pa}$. This planetary blackbody model was fit to the common-mode reduced spectra excluding the wavelength bins shown in this figure for each orbit at WFC3's native resolution. The black point in eclipse denotes the uncertainty on the in-eclipse flux measurement. In the common-mode method all data is divided by this in-eclipse data. Therefore, this uncertainty also represents the uncertainty on the mean deviation of each bin. We denote the mean deviation in the second visit with a horizontal line. The data for the first visit is faded because Section \ref{sec:analysis:phase} determined that the common-mode method is unreliable for the first visit. We plot the median planetary light-curve from Figure \ref{fig:rawdata} in aquamarine green in the background for reference of the timing of the peak in the excess flux. Its scale (right axis) is the same as for the binned data. The lower panel contains the corresponding stellar pulsation with the pulsation amplitude averaged over each orbit denoted as diamonds. Also denoted is the posterior distribution of the time of periastron from this work. The posterior median is demarcated with a vertical dashed line.\\
    \textbf{Right:} Wavelength binning used for the Pa$\beta$ line analysis in the left panel. The upper panel shows an example measured spectrum of a single WFC3 exposure. The dip in flux at 1.26\,$\mu$m is due to a dip in the G141 transmission function. The lower panel shows a PHOENIX stellar spectrum for HAT-P-2 \citep{Husser2013} smoothed to the WFC3 G141 dispersion resolution. The Pa$\beta$ line at 1.282\,$\mu$m is clearly discernible and has a band-averaged depth of ${\sim}5$\%. We collected the five WFC3 pixel-wide columns surrounding the Pa$\beta$ line into the dark red bin. The 6 neighboring pixel-wide columns were binned into the light blue bin and the 8 columns surrounding that were grouped into the outer, magenta bin. None of the bins contain any wavelengths of the other bins. 
    }
    \label{fig:paschenline}
\end{figure*}

To investigate the origin of this apparent excess flux, we first performed the common-mode method at the WFC3 native resolution, and found a flux excess in all five pixel-sized wavelength bins around 1.28\,$\mu$m. We therefore collected these five wavelength columns into a single bin. To compare this bin to the rest of the spectrum, we created two extra wavelength bins that surround the 1.28\,$\mu$m bin and that are unaffected by the stellar Pa$\beta$ line. The three bins are shown in the right panels of Figure \ref{fig:paschenline}. 

Similarly to Section \ref{sec:results:partialphase:blackbody} we fitted a blackbody through the common-mode spectra at the native WFC3 resolution, this time excluding the three wavelength bins discussed above. We subtracted the flux expected by the blackbody fit from the common-mode reduced spectra and we plot the deviation from this blackbody, $\delta_{\rm Pa}$, in the lower panel of Figure \ref{fig:paschenline}. Recall that, in order to find the common-mode reduced spectrum of an orbit, its spectrum has to be divided by the in-eclipse spectrum. Therefore, if the in-eclipse spectrum is anomalous by one standard deviation due to statistical scatter, all common-mode reduced spectra will appear anomalous in the opposite direction by the same value.
However, the mean of the Paschen line bin deviates from the blackbody model by 4.1\,$\sigma$, while the other two bins deviate by only $0.3\sigma$. The mean difference between the Paschen line bin and the other bins in the second visit is three times its measurement uncertainty. That difference is smaller in the first visit, but still at 1.8\,$\sigma$. The flux excess for the first visit may therefore possibly originate from strong noise in the in-eclipse spectrum. However, this still leaves the extra excess in the second visit unexplained. Also, the match with the atomic Pa$\beta$ line is unlikely to be coincidence.

The Paschen line bin does not contain any strong flat-field artefacts and there are no bright background stars. There also appear to be no abnormal detector defects, and there is no anomaly in the number of detected cosmic rays around $1.28$\,$\mu$m. 

A drift  of the spectrum on the detector in the wavelength direction in combination with the undersampling of the stellar Pa$\beta$ line can deteriorate the quality of WFC3 spectra \citep{Deming2013}. However, on an orbit-to-orbit basis there was very little drift in the wavelength direction during the second visit (see Figure \ref{fig:rawdata}b). Also, we negated this effect by centering any wavelength bin around the Pa$\beta$ line. This way, if the core of the line shifts slightly from one pixel to another, the line as a whole stays in the same bin and the total flux in that bin is therefore stable. 

Whereas there is almost no drift in the wavelength direction during the second visit, we observe a visit-long drift of 0.8\,pixels in the spatial-scan direction. The IR WFC3 flat-field is uncertain at a 1\% level. Given a scan-length of 400 pixels, we estimate that this drift may introduce a maximum uncertainty of $2\times0.01\times0.8/400 = 40$\,ppm per wavelength channel at the WFC3 native resolution. This is insufficient and it tends to average out over multiple pixels. 

Curiously, the maximum deviation in Figure \ref{fig:paschenline} occurs after periastron, but before the white light maximum of the planet. This suggests that there is a possible connection to the planet's orbit and that the response is quicker than the planetary atmospheric layers probed by the WFC3 continuum. We further discuss astrophysical origins for this flux excess in Section \ref{sec:discussion:Paschen}.

\subsection{Stellar contamination}
\label{sec:results:partialphase:stellar}
\begin{figure}
    \centering
    \includegraphics[width=\columnwidth]{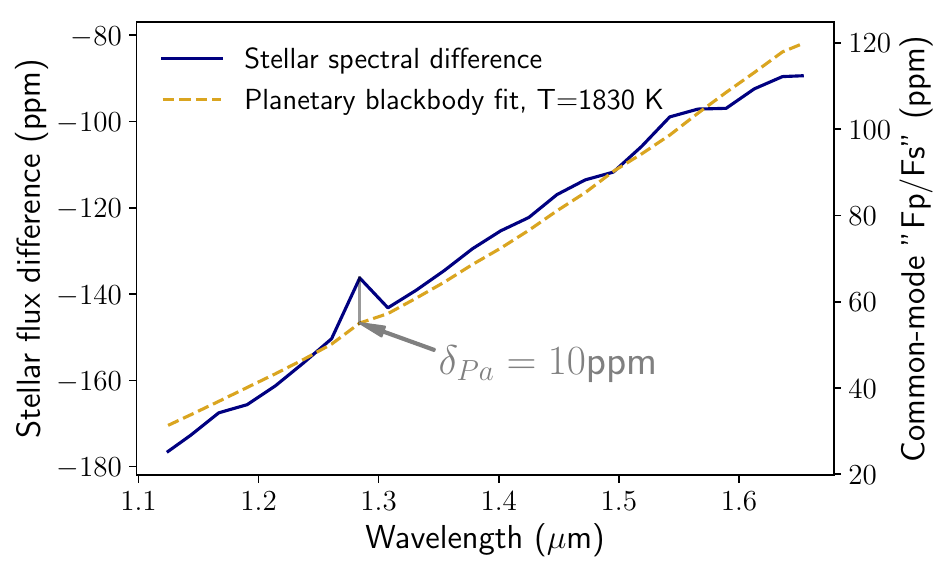}
    \caption{Spectral difference between two Kurucz stellar spectra of HAT-P-2 with a 0.5\,K difference in effective temperature. Suppose that the hotter stellar spectrum is the observed in-eclipse spectrum and that the cooler spectrum is the out-of-eclipse spectrum. If we perform a common-mode method on this, the resultant $F_p/F_s$ is shown on the right axis with the best-fit blackbody being a dashed line. Also, note that in this case the Paschen $\beta$ deviation (see Section \ref{sec:discussion:Paschen}) is measured to be 10\,ppm. All wavelength bins contain 5 wavelength pixel-columns. The $\delta_{\rm{Pa}}$ in this figure therefore corresponds to the dark red bin in Figure \ref{fig:paschenline}.
    }
    \label{fig:stellarmimic}
\end{figure}

A prerequisite for the correct application of the common-mode method is the absence of variations in the stellar spectrum. In this subsection we estimate the effects that a change in effective stellar temperature may have on the common-mode spectra. To do so, we used \texttt{pysynphot} \citep{pysynphot2013} and generated a Kurucz stellar spectrum \citep{kurucz1993} with stellar parameters closest to those observed for HAT-P-2 \citep{Tsantaki2014}.
We adopted this stellar spectrum as the in-eclipse spectrum. Subsequently, we generated another Kurucz spectrum, but with an effective temperature that is 0.5\,K lower and we assumed this to be an out-of-eclipse spectrum. For the sake of this argument, we set any planetary contribution to zero. Following the common-mode method stipulated by \citet{Arcangeli2021}, we divided the mock out-of-eclipse spectrum by the mock in-eclipse spectrum and retrieved the spectrum shown in blue in Figure \ref{fig:stellarmimic}. This surprisingly mimics the planetary spectra seen in Figure \ref{fig:phase}. Pretending that the spectrum in Figure \ref{fig:stellarmimic} was produced by a planet, we fitted a planetary blackbody to it plus a monochromatic offset and we retrieved a blackbody temperature of 1830\,K, displayed as a dashed, golden line. The mean difference between the two stellar spectra is approximately of the size of the constructively interfering stellar pulsations in white light. However, if stellar pulsations would truly change the effective stellar temperature in the way presented above, the star-induced slope in Figure \ref{fig:stellarmimic} would become negative, pushing for ''negative'' temperatures in Figure \ref{fig:GCMmodels}. This leaves open the possibility that the stellar effective temperature changed on a longer timescale throughout the second visit.

Also notable, is that a change in stellar temperature can induce a deviation from a planetary blackbody model like the deviation, $\delta_{\rm{Pa}}$ discussed in Section \ref{sec:results:Passchen}. However, for a difference in stellar effective temperature of 0.5\,K, this effect is on the order of just 10\,ppm, far smaller than the deviation observed in Figure \ref{fig:paschenline}. If the pulsations were the cause of the excess flux discussed in Section \ref{sec:results:Passchen}, the excess flux should correlate with the pulsations in the lower left panel of Figure \ref{fig:paschenline}. We only observe a Pearson correlation coefficient with $\delta_{\rm Pa}$ of -0.46 at a p-value of 0.06. There is also no significant Pa$\beta$ flux excess when comparing groups of exposures taken during a pulsation peak to groups of exposures taken during a pulsation trough.


Also, this effect is quantitatively not a satisfactory explanation for the flux excess discussed in Section \ref{sec:results:Passchen}, unless the stellar effective temperature changed by ${\sim}15$\,K throughout the visit and the telescope systematics counteracted this accordingly. However, such a large temperature change and hence large white-light variations would have been noticed before at other wavelengths \citep{Bakos2007, Lewis2013, deWit2017}.

The effect discussed in this section only affects common-mode reduced planetary spectra. In light-curve fitting methods, the stellar contamination would be filtered out with the visit-long polynomials or stellar pulsations.

\section{Discussion}
\label{sec:discussion}
The shape and timing of maximum and minimum planetary flux from the phase curve offer significant insights into HAT-P-2b's atmospheric properties, reflecting thermal variations resulting from time-variable heating. Comparing phase variations at different wavelengths provides further understanding of the atmosphere's thermal, wind, and chemical structure.

\subsection{Light-curve comparison to the forward models}
\label{sec:discussion:comparions_to_GCMs}


While the general shape and timing of the white light-curve is similar to that of the forward models, both underestimate the peak flux. The product of hydrogen dissociation, H$^{-}$, exhibits significant opacities at near-infrared wavelengths \citep{Arcangeli2018}, consequently reducing the pressures probed by WFC3. In the case of an inverted TP-profile, this would also increase the probed temperature at the peak of the light-curve. An overabundance of H$^{-}$ compared to the chemical equilibrium models may therefore explain the underestimated peak flux. Furthermore, the strong contrast between the flux increase timescale of $c_3=9.1^{+3.4}_{-2.1}$\,hr to the flux decrease timescale of $c_4=3.9^{+0.7}_{-0.6}$\,hr also indicates an enhanced H$^{-}$ abundance: as the planet heats up, H$^{-}$ is formed near-instantaneously in the hotspot \citep{BellCowan2018, Arcangeli2021}, prompting the formation of a thermal inversion, and decreasing the pressures probed by WFC3. Those layers have a shorter radiative timescale than the layers probed during the heating phase. This notion is corroborated by Figure \ref{fig:contrib_fct}: the best-fit 1D-retrieval probes higher altitude layers at shorter wavelengths than the forward models predict. Enhanced H$^{-}$ abundances are consistent with the 1D free-chemistry retrievals. This extra H$^{-}$ abundance with respect to the thermal equilibrium of models may originate from disequilibrium processes like photochemistry, as has been suggested for a couple other planets \citep{Lewis2020, Jacobs2022}. However, an increased H$^{-}$ opacity and therefore the probing of lower pressure layers, does not explain the big divergence between flux decrease timescale at $4.5$\,$\mu$m \citep[$10.3\pm1.5$\,hr;][]{deWit2017} and at WFC3 wavelengths \citep{Parmentier2018}. 

Alternatively, horizontal advection, and therefore wind speeds, may be higher than in the WFC3 photosphere than at the lower pressures probed by Spitzer, which is theoretically qualitatively in line with the GCM results by \citet{Lewis2014}, albeit not quantitatively. Stronger mixing decreases the radiative timescale for the planet as a whole because it is inversely proportional to the temperature cubed \citep{Iro2005, Seager2005}.

We measure a lag in the planetary peak brightness with respect to the incident stellar flux. This is partially due to the finiteness of the atmospheric radiative timescale \citep{Iro2010, CowanAgol2011}, and partially the result of the viewing geometry with more and more of the hotter dayside coming into view after periapse. 
This work's measurement of the peak flux delay of $c_2=6.7\pm0.6$\,hr is only slightly longer than the 1D time-stepping models (3-5 hours after periastron), the 3D GCMs (peak 4-5 hours after periastron), as well as the $4.5$\,$\mu$m Spitzer peak flux delay ($5.40\pm0.57$\,hr). 

The imperfect fit between the forward models and the fitted light-curve may also be the result of planetary variability. \citet{Lewis2014} measured the eclipse depth variability of HAT-P-2b in their GCMs without a thermal inversion. The planet's eclipse depth is likely variable by a maximum of ${\sim}20$\,\% at G141 wavelengths. 

\subsection{Interpretation of spectral retrievals}
The 1D free-chemistry retrievals strongly depend on the monochromatic offset. We tried to mitigate this by assuming a prior on this offset leveraged from the white light-curve fits. However, the common-mode method was devised by \citet{Arcangeli2021} to monitor partial phase-curves in a non-continuous way, inhibiting the use of light-curve fits. Without prior on the offset, the free-chemistry retrievals show an even wider array of possible atmospheric compositions. Additionally, the free-retrievals find significantly different adiabats below the radiative-convective boundary (Figure \ref{fig:TPs}). The rapid changes in irradiation that HAT-P-2b experiences are not expected to alter these deep layers in such a short time frame. The retrievals would be more robust if thermo-chemical equilibrium retrievals, which require fewer fit parameters, were used for common-mode reduced spectra \citep[see e.g.][]{Arcangeli2018, Coulombe2023}. However, those models are \textit{a priori} in an instantaneous equilibrium, while eccentric planets are not. As such, equilibrium models are less sensitive to the thermo-chemical processes taking place during the heating/cooling of the planet, and they may present less accurate results.

\citet{Changeat2022} retrieved a secondary eclipse spectrum from the same WFC3 data as we analyzed in this work (for a detailed comparison in methodology, see Appendix \ref{app:Changeat_Edwards}). Their spectrum is most consistent with a blackbody spectrum, although their 1D free-chemistry retrieval found some tentative evidence ($\mathcal{B}<1$) for a non-inverted TP-profile with H$_2$O and VO in absorption. Those results are consistent with the findings of this work within the margins of uncertainty.

\subsection{An isothermal atmosphere?}
One of the key questions surrounding HAT-P-2b discussed in previous works was the possible presence of a (transient) thermal inversion in HAT-P-2b's atmosphere \citep{Lewis2013, Mayorga2021}. We tested this using 1D free-chemical retrievals in Sections \ref{sec:results:retrieval} and \ref{sec:results:partialphase:retrievals}. We found a thermal inversion that disappears after periapse. The transient thermal inversions predicted by the 1D and 3D forward models indeed commence before periapse, but they also persist well into secondary eclipse. 

The tentative thermal inversion before periapse is driven by the 2.3\,$\sigma$ detection of H$^{-}$ bound-free absorption. However, H$^{-}$ could not be significantly detected in any subsequent spectra. Also, its retrieved volume mixing ratio of $-4.2^{+0.4}_{-0.5}$ is 3-4 orders of magnitude higher than thermo-chemical equilibrium models predict for KELT-9b \citep{Jacobs2022} and WASP-18b \citep{Coulombe2023}, both of which are hotter than HAT-P-2b. The retrievals also find the H$^{-}$ at much higher altitudes, where it should have an even lower abundance due to the mutual neutralization with positive ions \citep{Lothringer2018}.

Alternative fits to the before-periapse spectrum and the spectral retrievals at later phases, result in unrealistically high abundances of TiO at low significances. The WFC3 secondary eclipse spectrum is excellently fit with a blackbody. HAT-P-2b's near-infrared spectrum is therefore most likely near-featureless from periastron to eclipse. This would mean that there is an extra slope in the phase-resolved spectra that the common-mode method could not account for and that leads to a spurious almost-detection of H$^{-}$ before periapse. A possible source of the extra slope is small variations in the stellar effective temperature (Section \ref{sec:results:partialphase:stellar}). While the slopes of the common-mode-retrieved spectra may be unreliable, this is not necessarily true for spectral features obtained through this method.

The retrievals are one dimensional, while the planet is intrinsically three-dimensional. The equilibrium GCMs predict a local thermal inversion in just the planet's hotspot and a non-inverted TP-profile in the surrounding region on the dayside. The combined modeled spectrum shows a thermal inversion. However, the retrieved median TP-profiles after periapse are nearly isothermal. The individual solutions for each spectrum are less isothermal, but averaged over all solutions combine into near-isothermals. This shows that there are multiple possible solutions to each spectrum and that the actual emission spectra may be a combination of several distinct spectra emerging from chemically distinct areas of the dayside. Applying 1-dimensional retrievals to an intrinsically 2D/3D planet can lead to biased results and spurious detections of elements \citep{Taylor2020}.
However, a full 3D thermo-chemical equilibrium retrieval \citep{Challener2022} is outside the scope of this paper and the data precision does not warrant such a complex model. 

By the time of secondary eclipse, the Earth-facing hemisphere-averaged thermal structure of the atmosphere has not been fully inverted yet as predicted by the forward models. This is at odds with the predictions of forward models. The chemistry in the forward models responds to the changing temperature instantaneously, while in reality it might take time. However, as shown by \citet{Parmentier2018, Arcangeli2021}, the dissociation/recombination timescales of H$_2$ and TiO, the prime agent for thermal inversions, are on the order of seconds. Therefore, there must be other processes at play, like advection, that prevent the swift emergence of a strong thermal inversion. Additionally, the onset of rapid dissociation would help transition the hotspot of the planet from a low-ionization to a high-ionization state. This may result in the coupling of the gas flow in the hotspot to the local magnetic field. This could induce excess heating \citep{Helling2021}, and control large-scale dynamics in and around the ionized region \citep{RogersKomacek2014, Beltz2022}. This could therefore also impact the timescale over which thermal inversions could develop. 

It is unlikely that the blackbody-like spectra are produced by lofty clouds as clouds can considerably cool down a planet, which is not observed here. The sedimentation efficiency chosen for the cloudy forward models ($f_{\rm{sed}}=0.5$) is already low for a high-gravity planet like HAT-P-2 \citep{Saumon2008, Gao2018}.

\subsection{Paschen $\beta$ line}
\label{sec:discussion:Paschen}
In Section \ref{sec:results:Passchen} we observed a flux excess in a wavelength bin that coincides with the Paschen $\beta$ line, the most prominent stellar line at G141 wavelengths. It is persistent throughout HAT-P-2b's phase curve and it peaks between periastron and the peak of the white light-curve. This feature was found with the common-mode method, but it is also visible in the \texttt{RECTE}-reduced eclipse spectrum as well as in the eclipse spectrum derived by \citet{Changeat2022}. We assessed that this spectral feature is unlikely to be noise, unlikely to be the result of a change in stellar effective temperature, and it is unlikely to be of telescopic origin. In this section we briefly discuss other possible astrophysical origins.

As HAT-P-2b approaches its host star, its upper atmosphere is shock heated. Due to the short radiative timescales, the upper atmosphere experiences the most rapid changes and its temperature will peak before the lower atmosphere's temperature peaks \citep{Lewis2014}. Atomic/molecular lines probing upper layers therefore peak earlier than lines probing deeper layers. 

Studies on hotter exoplanets found that non-local thermal equilibrium (NLTE) effects in the upper atmosphere can significantly heat up the planetary upper atmosphere as well as alter the populations of hydrogen electron energy levels \citep{Wyttenbach2020}. NLTE effects therefore increase the strength of some planetary hydrogen lines in the transmission spectrum \citep{Fossati2021, Borsa2022}, which may include Pa$\beta$. However, NLTE effects on specific hydrogen lines have only been found in high-resolution spectra as they probe higher atmospheric layers than emission spectroscopy.

\citet{SanchezLopez2022} found planetary Pa$\beta$ absorption in KELT-9b's transit. They interpret their detection as hydrogen absorption from atmospheric layers at ${\sim}1.3$\,$R_P$. As an alternative, they entertain the thought that the absorption may originate from hydrogen atoms escaping the planet after stellar activity. This presents an analogy to HAT-P-2b's shock heating during periapse.

Alternatively, the signal in the 1.28\,$\mu$m bin could be induced by the planet on the host star. Star-planet interactions can have many forms and they have been observed on multiple planets \citep{Shkolnik2008, Desert2011b, Vidotto2020, Bryan2024}, including the pulsations on HAT-P-2 \citep{deWit2017}. We did not observe a significant stellar dimming corresponding to ${>}2$\,K. Hence, if the Pa$\beta$ signal is indeed induced by the planet, the planet's effect on that line must be greater than on the stellar continuum. The stellar Paschen line originates from higher altitudes in the stellar atmosphere than the continuum. It is therefore not inconceivable that only these higher layers temporarily either heat up or expand due to either external effects or the stellar pulsations, although $\delta_{\rm{Pa}}$ is only weakly correlated with the pulsations.

The interplay between the planetary magnetic field and the stellar magnetic field is found to induce chromospheric activity in planetary host stars \citep{Cauley2019}. \citet{Cuntz2000} propose that orbiting planets induce magnetic reconnecting events between the planetary and stellar magnetospheres which could lead to magnetic heating. 
Chromospheric activity is typically monitored through variations in the Ca {\sc II} H\&K line strength, but the Pa$\beta$ line is sometimes also used as an activity indicator \citep{Schmidt2012, Fuhrmeister2023}. 
Being a massive planet, HAT-P-2b likely has a strong magnetic field \citep{Yadav2017,Arcangeli2019}. 
Magnetic field line interactions are inversely proportional to the squared distance between the star and planet \citep{Cuntz2000}, which means that chromospheric activity should be strongly biased to the time the HAT-P-2b is closest to its host star. \citet{Klein2021} found correlations between the rotation of the magnetic pole of AU Mic with the strength of the stellar Pa$\beta$ line.

However, this does not very well explain that the 1.28\,$\mu$m bin has a clear minimum during eclipse. Also, whereas the flares on AU Mic could originate from star-planet interacations \citep{Ilin2022}, no multi-wavelength flare events have yet been discovered for HAT-P-2 \citep{Ilin2024}. 

Alternatively, the stellar heating may be caused by the Poynting flux communicating magnetic field energy from the planet to the star for sub-Alfvénic interaction, which is analogical to Io's or Ganymede's auroral footprint on Jupiter \citep{Saur2013, Hue2023, Sulaiman2023}. 
If the planet's Alfvén Mach number becomes larger than one (which occurs at 0.08 AU around the Sun), the magnetic interaction drops steeply, which may be the case at HAT-P-2b's apastron (0.1\,AU).
However, \citet{Figueira2016} were unable to find any correlations between the planetary distance and stellar activity indicators for the even more eccentric HD 80606b, which also orbits a quiescent star.

\section{Summary}

In this work, we investigated the atmospheric dynamics and chemistry of the highly-eccentric close-in giant HAT-P-2b by spectroscopically measuring its partial phase curve using WFC3. We fitted the white light-curve using an asymmetric Lorentzian and found the peak flux to occur $c_2=6.7\pm0.6$\,hr after periastron, ${\sim}2\sigma$ longer than forward model predictions and previous observations. We find a relatively large difference between the flux increase ($c_3=9.1^{+3.4}_{-2.1}$\,hr) and decrease ($c_4=3.9^{+0.7}_{-0.6}$\,hr) timescales. We explain this as well as the delayed peak flux with the possible emergence of H$^{-}$ opacities during periapse passage as higher altitudes are probed with shorter radiative timescales. Furthermore, we found that the observed geometry of the system has evolved since the Spitzer observations almost a decade earlier, but that the orbital harmonics stellar pulsations have likely remained remarkably stable. Mid-transit time took place at $t_0=2459204.11219\pm0.00014$ BJD TDB, which is discrepant with previous measurements \citep{Ivshina2022} at $3.3$\,$\sigma$, and the argument of periastron changed from $\omega_{\star}=188.44\pm0.43$\,degrees in 2011 to $\omega_{\star}=191.6^{+1.9}_{-2.0}$\,degrees in December 2020. We measure the time of periastron to take place $15.9\pm0.5$\,hr after mid-transit, which is $0.67\pm0.5$\,hr later than in the epoch observed by \citet{deWit2017} but consistent with the predictions from \citet{deBeurs2023} for the current observation epoch (December 2020).

We searched for the planet-induced stellar pulsations found by \citet{deWit2017} by fitting for two stellar pulsation modes at 79 and 91 times the orbital frequency of the planet. Indeed, we retrieve these pulsations at high significance ($\Delta$BIC=21) and at amplitudes similar to those found at $4.5$\,$\mu$m. Because of their small, seemingly monochromatic amplitudes, they should have little effect on the planetary climate and spectral continuum.

The measured transmission spectrum is flat as expected because of the computed small scale height for this planet, allowing us to provide a refined measurement of the planet's radius: $R_p^2/R_s^2 = 0.004869 \pm 0.000012$. The secondary eclipse spectrum is very well fit by a blackbody spectrum, and is therefore consistent with a mostly isothermal TP-profile. No molecules could be detected with high significance (${>}3\sigma$).

We applied the common-mode method by \citet{Arcangeli2021} to a partial WFC3 phase curve of HAT-P-2b with mixed results. The first visit was visibly too affected by telescope systematics. The second visit was less affected by instrumental effects, but the resulting spectra have a different spectral slope than expected. We show that small changes in stellar effective temperature can have outsized effects on the spectral slope measured by the common-mode method. Both the blackbody fits as well as the free-retrievals that used priors from the white light-curve fits, show inadequate results stemming from the monochromatic offset introduced by the common-mode method. This translates into blackbody fits with higher temperatures than expected and deep atmosphere adiabats that change unphysically through time. Future partial phase curves with HST will face the same issues. 

However, the common-mode reduced spectra of the partial phase curve revealed an anomalously high flux in the spectroscopic bin that coincides with the hydrogen Paschen $\beta$ line and is irrespective of spectral slope or monochromatic offset. On average, it deviates by $4.1$\,$\sigma$ from a planetary blackbody model over the duration of the second visit. The flux excess is unlikely to be of instrumental origin and it is not affected by the above mentioned challenges for common-mode reduced spectra. It is therefore likely of astrophysical origin. It does not seem to be correlated to stellar pulsations, but a stellar origin is not excluded. Ground-based emission spectroscopy focusing on additional hydrogen lines and potentially the helium line at $1.0833$\,$\mu$m could elucidate the source of the flux excess in the Pa$\beta$ bin.

\vspace{3em}
We thank L. Kreidberg for their contribution to the proposal that led to this paper, and we thank V. Panwar and H. Shivkumar for the interesting discussions.
J.M.D acknowledges support from the Amsterdam Academic Alliance (AAA) Program, and the European Research Council (ERC) European Union’s Horizon 2020 research and innovation program (grant agreement no. 679633; Exo-Atmos). This work is part of the research program VIDI New Frontiers in Exoplanetary Climatology with project number 614.001.601, which is (partly) financed by the Dutch Research Council (NWO). Support for program HST-GO-15131 was provided by NASA through a grant from the Space Telescope Science Institute, which is operated by the Associations of Universities for Research in Astronomy, Incorporated, under NASA contract NAS5-26555. Z.L.D. would like to thank the generous support of the MIT Presidential Fellowship, the MIT Collamore-Rogers Fellowship and to acknowledge that this material is based upon work supported by the National Science Foundation Graduate Research Fellowship under Grant No. 1745302. Z.L.D. acknowledges support from the D.17 Extreme Precision Radial Velocity Foundation Science program under NASA grant 80NSSC22K0848.
\vspace{5mm}
\facilities{HST (WFC3), Spitzer (IRAC)}


\software{\texttt{astropy} \citep{astropy2013},  
          \texttt{batman}, \citep{batman2015}
          \texttt{emcee}, \citep{emcee}
          \texttt{exoCTK} \citep{Bourque2021},
          \texttt{lmfit} \citep{Newville2014},
          \texttt{matplotlib} \citep{Hunter2007},
          \texttt{pandas} \citep{pandas2020},
          \texttt{POSEIDON} \citep{MacDonald2017, MacDonald2023},
          \texttt{PyMSG} \citep{Townsend2023},
          \texttt{PyMultiNest} \citep{Feroz2009, Buchner2014},
          \texttt{pysynphot} \citep{pysynphot2013},
          Python \citep{python1995},
          \texttt{scipy} \citep{scipy2020},
          \texttt{seaborn} \citep{Waskom2021},
          \texttt{RECTE} \citep{Zhou2017},
          }

\appendix

\section{Spectrum comparison to previous studies}
\label{app:Changeat_Edwards}

\begin{figure}
    \centering
    \includegraphics[width=0.5\columnwidth]{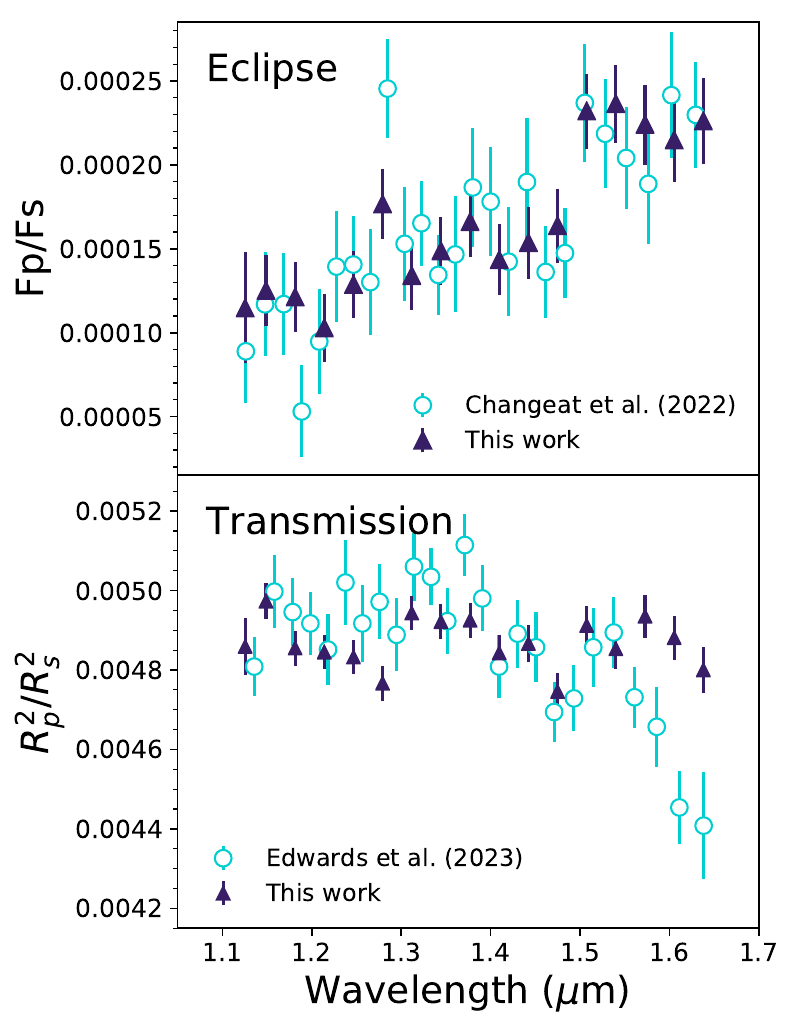}
    \caption{Comparison of the secondary eclipse and transmission spectra from previous studies to the ones presented in this study. Purple spectra denote spectra from this work and cyan spectra are from \citet{Changeat2022} and \citet{Edwards2022}.}
    \label{fig:compare_CE}
\end{figure}
\citet{Changeat2022} and \citet{Edwards2022} recently observed HAT-P-2b's secondary eclipse spectrum and transmission spectrum respectively. They used the same data presented in this work, but they used Iraclis \citep{Tsiaras2016a} to reduce the data and they used an exponential-ramp method \citep[e.g.][]{Kreidberg2014} to fit for the telescope systematics in the white light-curve. To find the transmission and eclipse spectra, they used the same orbits as we used in this work, but they removed the first exposure of each orbit. They divided each spectroscopic light-curve by the telescope systematics model derived from the white light-curve \citep[divide-white method;][]{Kreidberg2014} and then fit a secondary eclipse/transit model to the detrended light-curve. 

We compare the spectra from \citet{Changeat2022} and \citet{Edwards2022} to the spectra from this work in Figure \ref{fig:compare_CE}. Their uncertainties on each data point are 40\% and 80\% larger than the uncertainties presented in this work for the eclipse and transit spectra respectively. This difference is partially explained by them splitting the light into 50\% more wavelength bins and discarding every first exposure of every orbit. Also, they conducted separate fits for the eclipse and transit events, whereas our approach involves simultaneous fitting, informing the fits of one event with the orbit-long ramp shape from the other.

\citet{Changeat2022} fit for a planetary emission model constant with time, while we used an asymmetric Lorentzian phase curve model, which has a decreasing planetary flux around eclipse. If we had followed their emission model, the retrieved eclipse spectrum would have been ${\sim}45$\,ppm\,$\mu$m$^{-1}$ steeper. This is largely mitigated by the difference in light-curve fitting methods:
while the ``divide-white'' light-curve fitting method corrects wavelength-independent flux changes effectively \citep[see e.g.][]{Kreidberg2014, Barat2023}, it tends to underfit wavelength-dependent systematics, notably HST's orbit-long ramps. This wavelength dependence of the orbit-long ramps is especially evident when the product of the stellar spectrum with the G141 transmission function is strongly wavelength-dependent \citep{Kreidberg2014, Jacobs2022}. This is the case for HAT-P-2, introducing an extra negative slope of ${\sim}{-}20$\,ppm\,$\mu$m$^{-1}$ into the spectrum of \citet{Changeat2022}. 
The remaining ${\sim}-25$\,ppm\,$\mu$m$^{-1}$ is unexplained for, but well within the measurement uncertainties of $60$\,ppm\,$\mu$m$^{-1}$.

Another notable difference is that the eclipse depth of the bin centered at 1.28\,$\mu$m is much deeper for \citet{Changeat2022}. This is mainly caused  by the choice of binning: \citet{Changeat2022} chose narrower bins, such that their 1.28\,$\mu$m bin solely includes light from the Pa$\beta$ line. This is in contrast to the bin from this work, which also contains two pixel-wide columns at wavelengths shorter than 1.277\,$\mu$m. Those two columns do not show the Paschen anomaly discussed in Section \ref{sec:results:Passchen}. Another reason why the 1.28\,$\mu$m bin is more anomalous in the \citet{Changeat2022} spectrum, is that the Pa$\beta$ bin shows a stronger planetary flux decrease around eclipse than the surrounding bins (see Figure \ref{fig:paschenline}). The difference in eclipse depth because of the assumed planetary light-curve is therefore greater than in the surrounding bins.

The transit depths found by \citet{Edwards2022} decrease with wavelengths, whereas the spectrum found by this work is flat.
We found that we could reproduce the downward slope observed by \citet{Edwards2022} by using the same ramp amplitude parameter for all orbits. A single shared ramp amplitude is not common-practice \citep[see e.g.][]{Kreidberg2014} as the first non-discarded orbit is often fit with a different ramp amplitude than the rest. The WFC3 ramp amplitudes depend on the incoming flux and therefore also on wavelength \citep{Zhou2017}. As shown in Figure 1 of \citet{Jacobs2022}, the difference in ramp amplitude between the first non-discarded WFC3 orbit and the other orbits is therefore wavelength dependent. By inadequately fitting the ramp amplitude of the single orbit before transit, \citet{Edwards2022} therefore introduced a spurious slope to the transmission spectrum. 
\citet{Changeat2022} used the same exponential-ramp model as \citet{Edwards2022} to fit their secondary eclipse data. However, their emission spectrum is not biased by the single-ramp-amplitude assumption because the secondary eclipse measurements were preceded by more than a dozen orbits with the same ramp amplitude.

\section{Additional tables and figures}
\label{app:additional_figs}

\begin{figure*}
    \centering
    \includegraphics[width=\textwidth]{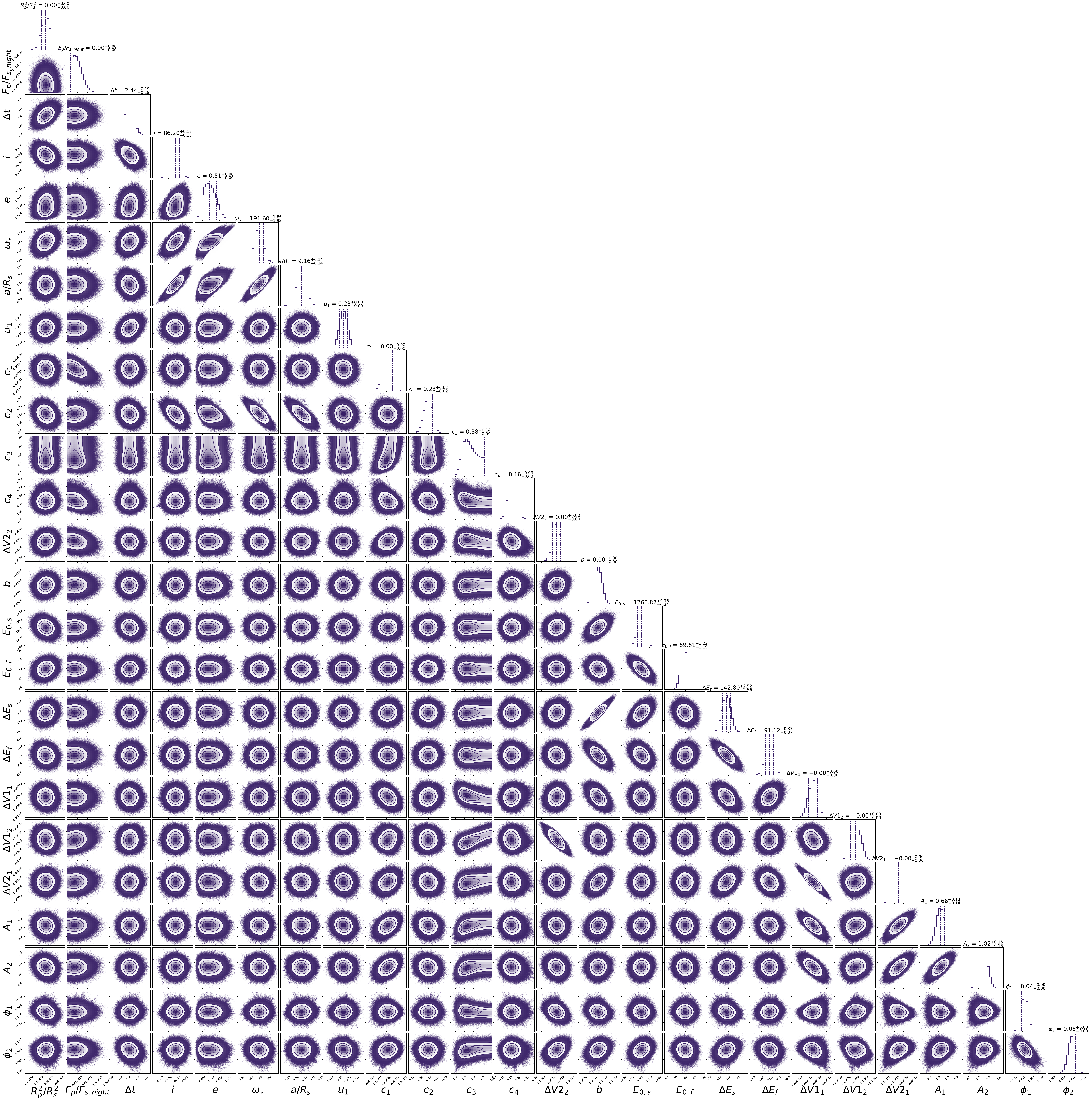}
    \caption{Corner-corner plot of the parameters retrieved from HAT-P-2b's white light-curve.
    }

    \label{fig:cornercorner}
\end{figure*}

\begin{table}[]
\begin{tabular}{llll}
Wavelength ($\mu$m) & $F_p/F_s$ (ppm) & $R_p^2/R_s^2$ & Reference \\
\hline
\hline
0.8 & - & $0.0047748\pm0.00006$ & \citet{Patel2022} \\
1.126 & $115_{-33}^{+34}$ & $0.00486\pm0.00008$ & This work \\
1.149 & $125\pm22$ & $0.00497\pm0.00005$ & This work \\
1.181 & $121\pm21$ & $0.00486\pm0.00005$ & This work \\
1.214 & $103\pm21$ & $0.00484\pm0.00005$ & This work \\
1.247 & $129_{-20}^{+21}$ & $0.00483\pm0.00005$ & This work \\
1.279 & $177\pm21$ & $0.00477\pm0.00005$ & This work \\
1.312 & $134\pm21$ & $0.00494\pm0.00005$ & This work \\
1.344 & $149\pm21$ & $0.00492\pm0.00005$ & This work \\
1.377 & $166\pm21$ & $0.00493\pm0.00005$ & This work \\
1.410 & $143_{-21}^{+22}$ & $0.00484\pm0.00005$ & This work \\
1.442 & $154\pm22$ & $0.00487\pm0.00005$ & This work \\
1.475 & $164_{-23}^{+22}$ & $0.00475\pm0.00005$ & This work \\
1.507 & $232\pm23$ & $0.00491\pm0.00005$ & This work \\
1.540 & $237\pm24$ & $0.00485\pm0.00006$ & This work \\
1.573 & $224_{-25}^{+24}$ & $0.00494\pm0.00006$ & This work \\
1.605 & $215_{-26}^{+25}$ & $0.00488\pm0.00006$ & This work \\
1.638 & $226\pm26$ & $0.00480\pm0.00006$ & This work \\
3.6 & $996\pm72$ & $0.00465\pm0.00011$ & \citet{Lewis2013} \\
4.5 & $971\pm21$ & $0.00494\pm0.00004$ & \citet{deWit2017} \\
5.0 & $710_{-130}^{+290}$ & - & \citet{Lewis2013} \\
8.0 & $1392\pm95$ & $0.00446\pm0.00022$ & \citet{Lewis2013}
\\
\hline
WFC3- & & \multirow{2}{*}{$0.004869 \pm 0.000012$}  & 
\multirow{2}{*}{This work}\\
averaged
\end{tabular}
\caption{Measured eclipse depths and transit depths of HAT-P-2b at different wavelengths from this work and from the literature. We also provide the weighted wavelength-averaged transit depth of all WFC3 data. \citet{Patel2022} did not measure the eclipse depth and \citet{Lewis2013} did not take any data during primary transit at $5.0$\,$\mu$m.}
\label{app:tab:transit}
\end{table}

\begin{figure*}
    \centering
    \includegraphics[width=\textwidth]{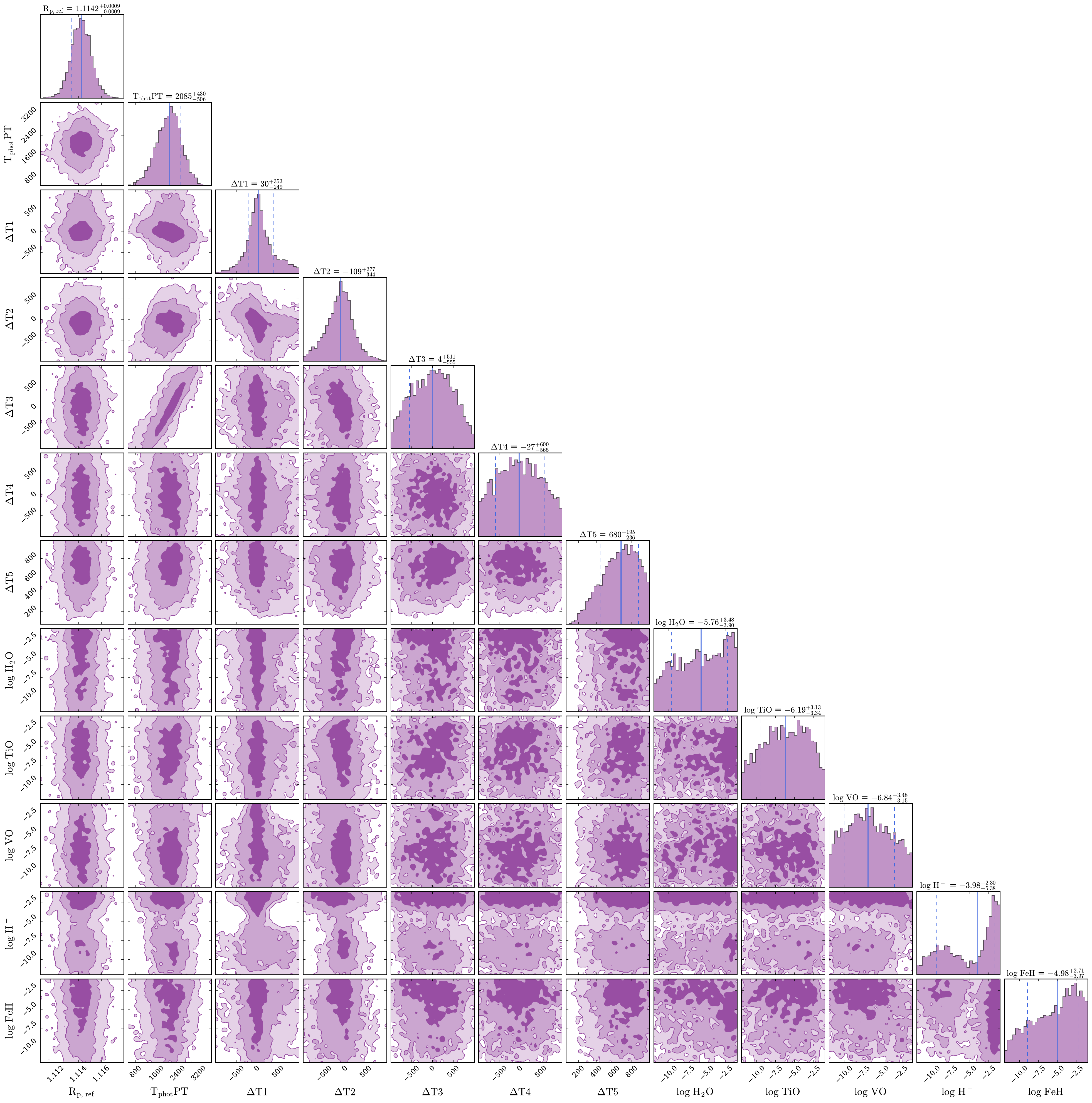}
    \caption{Corner-corner plot of the free-chemistry retrieval of the secondary eclipse spectrum. The reference radius at 0.01 bar ($\rm R_{p,ref}$) is in Jupiter radii, $\Delta {\rm T}_i$ are the temperature changes between the layers listed in Section \ref{sec:1D_free-chemistry_models}, and the abundances of the fitted molecular species are in log(volume mixing ratio).
    }

    \label{fig:eclipse_retrieval_corner}
\end{figure*}

\begin{figure*}
        \centering
            \includegraphics[width=\textwidth]{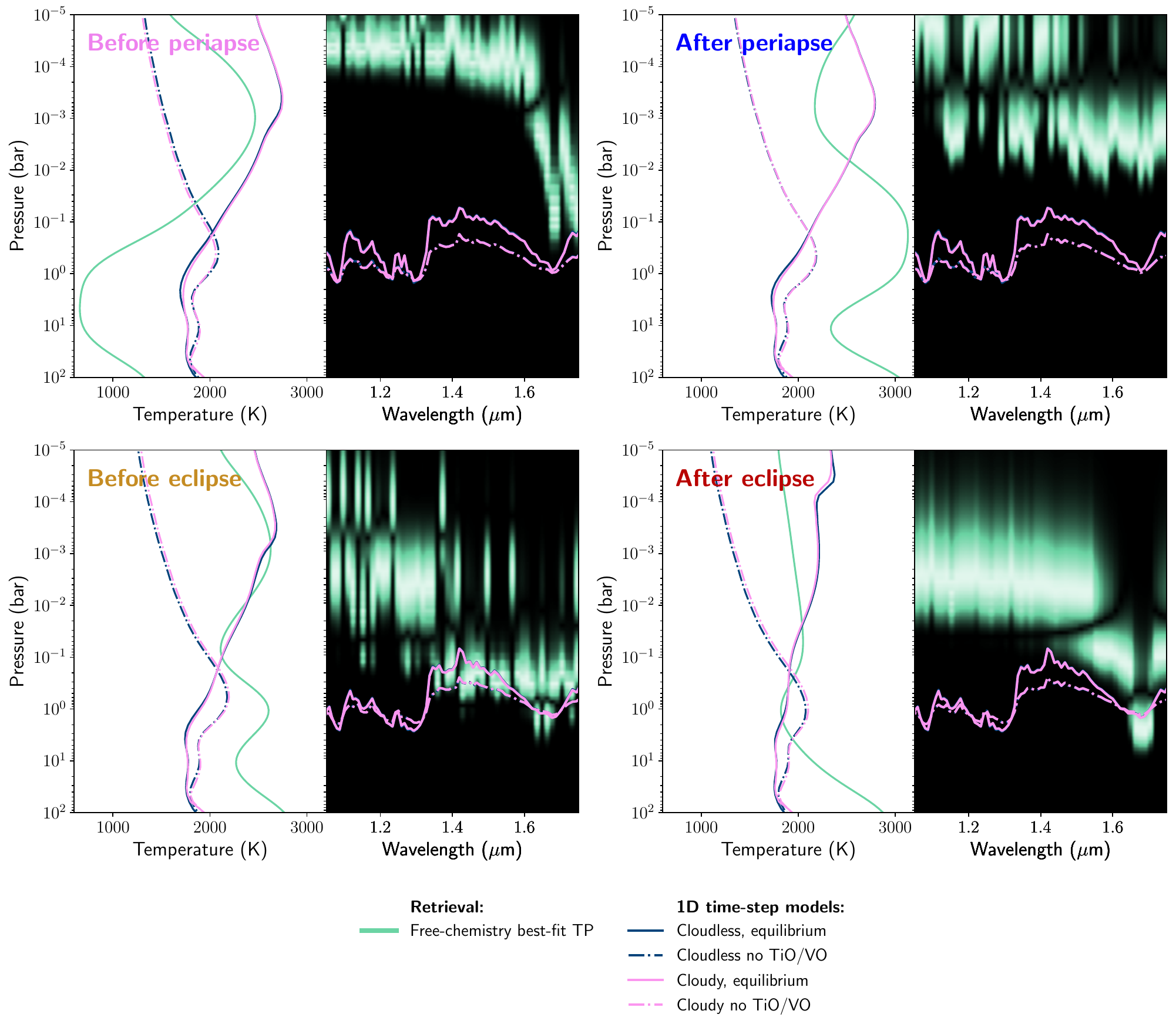}  
        \caption{Best-fit temperature pressure profiles (aquamarine green) of the 1D free-chemistry retrievals of the four phase-resolved emission spectra of HAT-P-2b with the corresponding contribution functions. As a reference, the TP-profiles and the $\tau=0.5$ contribution functions are shown for the 1D time-stepping forward models.} 
        \label{fig:contrib_fct_phase}
    \end{figure*}

\bibliography{references}{}
\bibliographystyle{aasjournal}

\end{document}